\documentclass[fleqn,11pt,DIV=13]{scrartcl}

\usepackage[english]{babel} 
\usepackage[ansinew]{inputenc} 
\usepackage{microtype}
\usepackage{multirow,
						graphicx,
						booktabs,
						upgreek
						}
\usepackage[bitstream-charter]{mathdesign}
\usepackage[margin=10pt,font=small,format=plain]{caption}
\usepackage{hyphenat}
\usepackage[fleqn]{amsmath}
\usepackage{cite}
\usepackage[
    pdftitle={ChiSRs},
    pdfauthor={Buchheim, Hilger, Kaempfer,Leupold}
    bookmarks=true,
    pdfborder={0 0 0},
    bookmarksopen=true,
    unicode,
    hyperfootnotes=true,
    colorlinks=false,
    pagebackref=false,
    citecolor=blue,
    linkcolor=blue,
    urlcolor=blue,
    breaklinks,
    pdfpagelabels,
    ]{hyperref}
\usepackage[numbered]{bookmark}
\usepackage[all]{hypcap}
\usepackage{doi}
\usepackage[
	toc,
	acronym,
	style=long,
	nonumberlist,
	nowarn,
	nomain,
	nopostdot
	]{glossaries}%
\makeglossaries
\renewcommand*{\CustomAcronymFields}{%
  name={\the\glsshorttok},
  description={\the\glslongtok},
  first={\noexpand\the\glslongtok\space(\the\glsshorttok)},%
  firstplural={\noexpand\the\glslongtok\noexpand\acrpluralsuffix\space(\the\glsshorttok\noexpand\acrpluralsuffix)},%
  text={\the\glsshorttok},%
  plural={\the\glsshorttok\noexpand\acrpluralsuffix}%
}
\SetCustomStyle
\glsdisablehyper
\newacronym{OPE}{OPE}{operator product expansion}
\newacronym{QSR}{QSR}{QCD sum rule}
\newacronym{DCSB}{D$\upchi$SB}{dynamical chiral symmetry breaking}
\newacronym{RGI}{RGI}{renormalization group invariant}

\newcommand{\rmd}{\mathrm{d}}

\newlength\imageheight

\begin{document}

\title{Chiral-partner D mesons in a heat bath within QCD sum rules}

\author{T.\ Buchheim${}^{1,2}$, T.\ Hilger${}^{3,4}$, B.\ K\"{a}mpfer${}^{1,2}$, S.\ Leupold${}^{5}$}

\date{
						\small{
						${}^{1}$ Helmholtz-Zentrum Dresden-Rossendorf, PF 510119, D-01314 Dresden, Germany\\
						${}^{2}$ Technische Universit\"{a}t Dresden, Institut f{\"u}r Theoretische Physik, D-01062 Dresden, Germany\\
						${}^{3}$ Institute of High Energy Physics, Austrian Academy of Sciences, A-1050 Vienna, Austria\\
						${}^{4}$ University of Graz, Institute of Physics, NAWI Graz, A-8010 Graz, Austria\\
						${}^{5}$ Uppsala Universitet, Institutionen f{\"o}r fysik och astronomi, Box 516, 75120 Uppsala, Sweden\\[3ex]
						}
			}

\publishers{\small{\today}}

\maketitle

\begin{abstract}
Utilizing {QCD} sum rules, we extract the temperature dependences of the spectral properties of the pseudo-scalar and scalar D mesons regarded as chiral partners.
Besides the masses also decay constants are analyzed as the D meson yields in heavy-ion collisions may be sensitive to their altered decay properties in an ambient strongly interacting medium.
Our findings are ($i$) a decreasing scalar D meson mass for growing temperatures while its pseudo-scalar partner meson seems hardly affected, which is in qualitative agreement with hadronic model calculations; ($ii$) inferring an equally weak temperature dependence of the pseudo-scalar D meson decay properties the decreasing residua and decay constants of the scalar particle point towards partial chiral restoration.
As a bonus of our analysis in the pseudo-scalar sector we determine the pseudo-scalar decay constant at vanishing temperature.
Due to the connection to particular leptonic branching fractions this decay constant is of great interest allowing for the determination of the off-diagonal CKM matrix element $|V_{cd}|$ at zero temperature.
\end{abstract}

\section{Introduction}
\label{sec:intro}
\label{sec:MotivChSRs}

Open charm mesons uncover various fundamental features of {QCD}.
In relation to light quarks ($q$), the remnant of spontaneously broken chiral symmetry \cite{Wise:1992hn} should show up, while the heavy quarks ($Q$) give rise to another special symmetry resulting in the heavy-quark expansion \cite{Neubert:1993mb}.
That is, the open charm mesons feature via $m_q < \Lambda_\mathrm{QCD} < m_Q$ the hierarchy of masses $m_{q,Q}$ w.\,r.\,t.\ the scale $\Lambda_\mathrm{QCD}$ \cite{Olive:2016xmw} emerging from dimensional transmutation together with the specifics of light and heavy quark dynamics formally tied to the expansions of $m_q \rightarrow 0$ and $m_Q \rightarrow \infty$.
The center symmetry, essentially related to confinement, in contrast, is anchored in the flavor-blind glue dynamics \cite{Greiner:2002ui}.
Owing to the relation $m_q \ll m_Q$, open charm (or even better, bottom) mesons can be considered as two-body  bound states with valence quark structure $\bar q Q$ or $\bar Q q$.
Thus, one can consider open charm mesons as QCD-type hydrogen atoms, though, the current quark mass of the light quark makes the system relativistic.
In heavy-ion collisions at high beam energies, e.\,g.\ at {LHC} and {RHIC}, charm degrees of freedom enjoy some abundancy \cite{Adam:2015sza,Leitch:2006ff,Scomparin:2016gog,Plumari:2017ntm,Andronic:2017pug}, such that they can serve as probes of the above quoted facets of {QCD}: chiral symmetry breaking and its restoration, confinement and deconfinement, charm-hadron spectroscopy etc.
Further interesting aspects of open flavor mesons are rare CP-violating decays \cite{Beringer:1900zz} and their role w.\,r.\,t.\ exotic quantum numbers \cite{Hilger:2016drj}.
In this paper, we utilize the \gls{QSR} tool \cite{Shifman:1978bx} and compare chiral-partner open-charm mesons, that is the pseudo-scalar ($\mathrm{P}$) D meson and the scalar ($\mathrm{S}$) D meson.

\subsection{Mass modifications}

Chiral partners in the light-quark sector, e.\,g.\ the spectra of $\uprho$ and $a_1$ mesons, should become mass degenerate under chiral restoration \cite{Kapusta:1993hq,Hohler:2013eba}, but also the individual fate of a member in the chiral multiplet under chiral restoration is of interest \cite{Hilger:2010cn,UweHilger:2012uua,Hilger:2015zva}.
As mentioned above, also the open-charm mesons are subject to chiral restoration, albeit in a different manner.

Studying patterns of \gls{DCSB} with mesons containing a heavy and a light quark seems to be superimposed by the explicit symmetry breaking due to the non-negligible heavy-quark mass.
However, the original light-quark problem can be translated into the heavy-light sector, if the chiral symmetry transformations are restricted to the light-quark content.
As exemplified in Appendix~\ref{app:chitrafo} for the $\mathrm{D}^0$ meson, the respective pseudo-scalar current can be transformed into the scalar current by a finite chiral transformation with a specific set of rotation parameters.
Thus, the pseudo-scalar and scalar $\mathrm{D}^0$ mesons are qualified as chiral partners, which would have degenerate spectral properties in a chirally symmetric world.
In contrast, the experimentally verified masses \cite{Olive:2016xmw},
$m_\mathrm{P} = 1.865\,\mathrm{GeV}$ and $m_\mathrm{S} = 2.318\,\mathrm{GeV}$, deviate by about $450\,\mathrm{MeV}$ (similar to the $a_1$-$\uprho$ mass splitting) signaling the dynamical breakdown of chiral symmetry which is driven by order parameters \cite{Hilger:2009knMisc,Hilger:2011cq,Hilger:2012db,Buchheim:2015xka}, e.\,g.\ most prominently by the chiral condensate $\langle \bar q q \rangle$ as well as by chirally odd four-quark condensates which vanish in a chiral restoration scenario.
The \gls{DCSB} phenomenology, e.\,g.\ the splitting of chiral partner masses, can also be inferred from Weinberg sum rules in the heavy-light sector \cite{Hilger:2010zb}.
While in the pure light-quark sector, the \gls{RGI} quantity $m_q\langle \bar q q \rangle$ is essentially the steering quantity directly related to chiral aspects, for open-charm mesons, the combination $m_Q\langle \bar q q \rangle$ has been identified \cite{Hilger:2008jg,Hilger:2011cq,Buchheim:2014rpa,Buchheim:2015xka} as a central quantity, with much larger numerical impact due to the large charm quark mass.
 
In line with \cite{Kapusta:1993hq,Holt:2012wr} one may discuss several patterns of approaching chiral restoration. Masses might change as $m_P \rightarrow m_S$ or $m_S \rightarrow m_P$ or both changing.
Also more complicated changes of the spectral properties are conceivable like mixing or a melting of the spectra as a precursor to deconfinement.
The hadronic approach in \cite{Sasaki:2014asa} favors a decreasing scalar D meson mass for growing temperatures, ultimately approaching the pseudo-scalar D meson mass which essentially remains at its vacuum value.
Astonishingly, a simple ratio \gls{QSR} analysis in the spirit of Ioffe \cite{Ioffe:1981kw}, where the continuum of the spectral function is removed and the Borel mass is taken as the charm quark mass, gives a comparable result.
Here the pseudo-scalar D meson mass mildly grows for rising temperatures from $m_\mathrm{P} = 1.945\,\mathrm{GeV}$ in vacuum to $1.997\,\mathrm{GeV}$ at $T=150\,\mathrm{MeV}$, while the scalar D meson mass drops somewhat stronger from $m_\mathrm{S} = 2.477\,\mathrm{GeV}$ to $2.339\,\mathrm{GeV}$.
For a mutual judging of these restoration patterns, it is the first goal of this paper to contrast the findings in \cite{Sasaki:2014asa} and the Ioffe estimates with results from rigorous finite-temperature \glspl{QSR} complementing our previous studies~\cite{Hilger:2008jg,Hilger:2011cq} at finite net-nucleon densities.

\subsection{Modifications of decay properties}

An often disputed aspect in previous investigations is whether in-medium modifications can be traced back to observables, thus providing a lever arm to quantify the above mentioned {QCD} features.
Analogous to strangeness degrees of freedom \cite{Hartnack:2011cn}, for instance, the question has been addressed whether a 'shift' of the effective hadron masses due to strong interactions with the ambient medium changes the yields of the respective hadron species.
Reference~\cite{Andronic:2007zu} gives a tentatively negative answer, unless extremely tiny effects would be accessible.
However, it should be recalled that even the seemingly innocent quest for a 'mass shift', e.\,g.\ for the pseudo-scalar D meson suffers from some controversy within the \gls{QSR} approach \cite{Hayashigaki:2000es,Hilger:2008jg,Azizi:2014bba,Wang:2015uya,Suzuki:2015est}.

Due to the meson mass malaise one may shift the focus onto the decay properties of the mesons.
Investigations of mesonic decay properties, e.\,g.\ widths and decay constants, have attracted much attention in recent years \cite{Bordes:2004vu,Bordes:2005wi,Lucha:2010ea,Lucha:2011zp,Lucha:2014xla,Gelhausen:2013wia,Narison:2012xy,Narison:2015nxh,Wang:2015mxa}, not only in the realm of \glspl{QSR} \cite{Fukaya:2016fzs,Fahy:2015xka,Durr:2016ulb}, because accurate numerical values of open-flavor meson decay constants could be used to constrain, inter alia, off-diagonal {CKM} matrix elements from weak decays of these mesons.
To this end, an optimized \gls{QSR} approach has been introduced~\cite{Lucha:2010ea} tailored to extract mesonic decay constants.
It is the second goal of this paper to deduce and contrast for the first time the temperature dependences of the D meson decay properties from the new and the conventional \gls{QSR} approach.
This widens the quest for chiral restoration patterns at finite temperatures beyond chiral partner meson mass degeneracy, thus contributing to the understanding of in-medium properties of chiral partner mesons. 

While vacuum spectral properties of the pseudo-scalar D mesons are experimentally well constrained, only limited information on scalar D mesons is currently available.
The \gls{QSR} investigations of these particular mesons exhibit a congruent pattern:
Pseudo-scalar D mesons attracted much attention, where recent studies focus either on precise predictions of spectral or {QCD} parameters in vacuum \cite{Narison:2015nxh} or on medium modifications \cite{Gubler:2015qok,Suzuki:2015est}; whereas scalar D meson \glspl{QSR} have been rarely analyzed so far, primarily in vacuum \cite{Hayashigaki:2004gq,Narison:2003td,Sungu:2010zz,Wang:2015mxa} or cold nuclear matter \cite{Hilger:2010zf}.
As the idea of heavy-light mesons as probes of \gls{DCSB} gains acceptance also further investigations of chiral partners based on different approaches are performed \cite{Harada:2003kt,Molina:2008nh,Sasaki:2014asa,Suenaga:2014sga,Park:2016xrw} emphasizing the role of the scalar D meson.
Accordingly, we set up the in-medium \gls{QSR} for the scalar D meson in order to seek for signals of chiral restoration at finite temperatures in a particle--anti-particle symmetric medium.\\

Our paper is organized as follows:
In Sec.~\ref{sec:TdepPS} we provide the necessary ingredients to formulate finite-temperature \glspl{QSR} of pseudo-scalar and scalar open charm mesons, while the numerical input is relegated to Sec.~\ref{sec:numinput}.
Section~\ref{subsec:convBA} accommodates the conventional evaluation of D meson sum rules at finite $T$, where we give insights into important intermediate results in order to understand the distinct temperature behaviors of pseudo-scalar and scalar D mesons in the scope of \glspl{QSR}.
These results, in particular the obtained decay properties, are confronted with a \gls{QSR} evaluation with given meson-mass input in Sec.~\ref{sec:BAgivInput}.
Thereby, we can extract the off-diagonal CKM matrix element $|V_{cd}|$ from the pseudo-scalar vacuum decay constant as well.
We summarize our findings in Sec.~\ref{sec:sum}.
Further details on chiral symmetry transformations in the heavy-light sector, temperature effects on the pseudo-scalar and scalar \glsentryplural{OPE} as well as on the \gls{QSR} approach for extracting decay constants are provided in the appendices.

\section{Finite-temperature sum rules of pseudo-scalar and scalar D mesons}
\label{sec:TdepPS}

As we aim for the temperature dependences of the masses, residua and decay constants of pseudo-scalar and scalar D mesons, finite-temperature ($T$) \glspl{QSR} are evaluated based on the thermal average ($\langle \cdots \rangle_T$) of the time-ordered ($\mathrm T$) current-current correlator \cite{Hatsuda:1992bv}
\begin{align}\label{eq:ccc}
	\Pi_X (p;T) = i\int \rmd^4 x \; e^{ipx} \langle \mathrm{T} [j_X(x) j_X^\dagger (0)] \rangle_T
\end{align}
with momentum $p$ and the interpolating current $j_X$ which couples to the desired meson species~$X$.
The obtained vacuum results can be compared directly to decay constants from recent \gls{QSR} analyses using the following currents:
\begin{subequations}\label{eq:PScurrents}
\begin{align}
	j_\mathrm{P}(x) & = \partial_\mu j^\mu_\mathrm{A}(x) = (m_Q + m_q) \bar q (x) i\gamma_5 Q(x) \, , \label{eq:Pcurrent}\\
	j_\mathrm{S}(x) & = i \partial_\mu j^\mu_\mathrm{V}(x) = (m_Q - m_q) \bar q (x) Q(x)
\end{align}
\end{subequations}
with the canonical definition of vector and axial-vector currents, i.\,e.\ $j^\mu_\mathrm{V}(x) = \bar q(x) \gamma^\mu Q(x)$ and $j^\mu_\mathrm{A}(x) = \bar q(x) \gamma^\mu \gamma_5 Q(x)$, both being part of weak $\mathrm{V}-\mathrm{A}$ currents.
As demonstrated in Eqs.~\eqref{eq:PScurrents} by employing the quark equations of motion, the pseudo-scalar and scalar currents overlap with $j^\mu_\mathrm{A}$ and $j^\mu_\mathrm{V}$, respectively, admitting a weak decay of the D mesons, e.\,g.\ $\mathrm{D}^+ \rightarrow W^{+\,*} \rightarrow \ell^+ \nu_\ell$ (see the review on leptonic decays of charged pseudo-scalar mesons in~\cite{Olive:2016xmw}).
The pseudo-scalar D meson decay constant $f_\mathrm{P}$ defined as
\begin{align}\label{eq:deff_Palaf_pi}
	\langle 0 | j^\mu_\mathrm{A}(x) | \mathrm{P}(\vec{p}) \rangle = i p^\mu f_\mathrm{P} e^{-ipx} \, ,
\end{align}
analogously to the pion decay constant, reappears in
\begin{align}\label{eq:deffPNarison}
	\langle 0 | j_\mathrm{P}(0) | \mathrm{P} \rangle = f_\mathrm{P} m_\mathrm{P}^2
\end{align}
by virtue of Eq.~\eqref{eq:Pcurrent}, where $| \mathrm{P} \rangle$ denotes the pseudo-scalar D meson state with mass $m_\mathrm{P}$.
Accordingly, the decay constant $f_\mathrm{P}$ contributes to the leptonic decay width of lepton flavor~$\ell$
\begin{align}\label{eq:width}
	\Gamma(\mathrm{P}^+ \rightarrow \ell^+ \nu_\ell) = \frac{G_\mathrm{F}^2}{8\pi}\, |V_{Qq}|^2\, f_\mathrm{P}^2\, m_\ell^2\, m_\mathrm{P} \left( 1 - \frac{m_\ell^2}{m_\mathrm{P}^2} \right)^2 \, ,
\end{align}
the numerical value of which can be obtained from the branching ratio $\Gamma(\mathrm{P}^+ \rightarrow \ell^+ \nu_\ell)/\Gamma_\mathrm{tot}$ in \cite{Olive:2016xmw}.
Analogous definitions and relations hold for the scalar particle $\mathrm{S}$, however, experimental results for the wanted branching fraction of a decay into a specific leptonic channel, $\Gamma(\mathrm{S}^+ \rightarrow \ell^+ \nu_\ell)/\Gamma_\mathrm{tot}$, are not yet available since $\Gamma_\mathrm{tot}$ is governed by the strong interaction \cite{Olive:2016xmw} which drives the branching ratio tiny.%
\footnote{Although, no charged scalar D meson $\mathrm{S}^+$ is listed in \cite{Olive:2016xmw} it is assumed in this work to coincide with the uncharged D meson $m_{\mathrm{S}^+} \simeq m_{\mathrm{S}^0}$ due to iso-spin symmetry. This is also in agreement with the small SU(3) light-quark flavor breaking~\cite{Narison:2003td}, i.\,e.\ $m_{\mathrm{S}^+_s} \simeq m_{\mathrm{S}^+}$, and supported by analogy to the pseudo-scalar channel $m_{\mathrm{P}^+} \simeq m_{\mathrm{P}^0}$.}

A dispersion relation, which links the correlator~\eqref{eq:ccc} to its imaginary part, allows for two representations of $\Pi_X$: ($i$) it can be expanded into an asymptotic series of thermally averaged, local {QCD} operators, which is termed \gls{OPE}, and ($ii$) it can be expressed by the spectral density $\rho_X = \mathrm{Im}\Pi_X/\pi$ encoding the phenomenological properties of the meson~$X$. 
In order to improve the reliability of such a sum rule a Borel transformation, $\Pi_X(p;T) \longrightarrow \widehat\Pi_X(M^2;T)$, is performed, where for mesons at rest, $p=(p_0,\vec 0)$, the dependence on the meson energy in the deep Euclidean limit  $p_0^2 \rightarrow -\infty$ is traded for a Borel mass dependence $M$ \cite{Hatsuda:1992bv}.
The Borel transformed in-medium dispersion relations as well as \gls{OPE} formulae for pseudo-scalar currents \cite{Hilger:2008jg} can be easily rewritten for scalar mesons, e.\,g.\ using Eq.~(3.3) in \cite{Hilger:2011cq} and Eq.~(3) in \cite{Buchheim:2015xka} for the sum rule pieces being even and odd in $p_0$ \cite{Jin:1992id}, respectively.
At finite temperature but zero net-baryon density, where D and anti-D meson properties are degenerate,%
\footnote{In a particle--anti-particle asymmetric medium, i.\,e.\ at finite net-baryon density, open flavor mesons feature meson--anti-meson mass splitting \cite{Hilger:2008jg} which considerably complicates the extraction of the corresponding residua, cf.\ Ref.~\cite{Kampfer:2010vk}.}
the \glspl{QSR} reduce to
\begin{align}\label{eq:QSR}
	\int_0\limits^\infty \rmd s \, e^{-s/M^2} \rho_\mathrm{P,S}(s;T) \tanh\left(\frac{\sqrt{s}}{2T}\right) = \widehat{\Pi}_\mathrm{P,S}(M^2;T)
\end{align}
with the finite temperature \gls{OPE} derived in the $\overline{\mathrm{MS}}$ scheme and in the 'light chiral limit', $m_q \rightarrow 0$,
\vspace{-0.1ex}
\begin{align}\label{eq:OPETdep}
	\widehat{\Pi}_\mathrm{P,S} (M^2;T) 
	& = \frac{1}{\pi} \int\limits_{m_Q^2}^\infty \mathrm{d}s \; e^{-s/M^2} \mathrm{Im} \Pi^\mathrm{pert}(s) + e^{-m_Q^2/M^2} m_Q^2 \Bigg( \mp m_Q \langle \bar q q \rangle_T + \frac{1}{12}\langle \frac{\alpha_\mathrm{s}}{\pi} G^2 \rangle_T \nonumber\\
	& \phantom{=} + \left[ \left( \frac{7}{18} + \frac{1}{3} \ln\frac{\mu^2 m_Q^2}{M^4} - \frac{2\gamma_\mathrm{E}}{3} \right) \left( \frac{m_Q^2}{M^2} - 1 \right) - \frac{2}{3}\frac{m_Q^2}{M^2} \right] \langle \frac{\alpha_\mathrm{s}}{\pi} \left( \frac{(vG)^2}{v^2} - \frac{G^2}{4} \right) \rangle_T \nonumber\\[0.5ex]
	& \phantom{=} + 2\left( \frac{m_Q^2}{M^2} - 1 \right) \langle q^\dagger i D_0 q \rangle_T \pm \frac{1}{2} \left( \frac{m_Q^3}{2M^4} - \frac{m_Q}{M^2} \right) \left( \langle \bar q g \sigma G q \rangle_T - \langle \Delta \rangle_T \right) \Bigg) \, ,
\end{align}
where the perturbative contribution $\mathrm{Im} \Pi^\mathrm{pert}(s)$ obtained from Cutkosky's cutting rules can be found, e.\,g.\ in Refs.~\cite{Aliev:1983ra,Zschocke:2011aa} (mind the interpolating currents in \cite{Aliev:1983ra,Zschocke:2011aa} which differ from Eq.~\eqref{eq:PScurrents} by the quark mass factors).
The quantity $\mu$ denotes here the renormalization scale, and $\gamma_\mathrm{E}$ is the Euler-Mascheroni constant.
The expectation values $\langle (\alpha_\mathrm{s}/\pi)[(vG)^2/v^2 - G^2/4]\rangle_T$, $\langle q^\dagger i D_0 q \rangle_T$ and $\langle \Delta \rangle_T = 8 \langle \bar q D_0^2 q \rangle_T - \langle \bar q g \sigma G q \rangle_T$ are medium-specific condensates \cite{Buchheim:2015yyc}, thus they vanish at zero temperature by definition.
Quantitatively, dimension-6 four-quark condensate terms do not significantly contribute to $\widehat{\Pi}_\mathrm{P}$ \cite{Buchheim:2014rpa}; the same holds true for the scalar \gls{OPE}.
Thus, we restrict the numerical evaluation to condensates up to mass dimension~5.

Henceforth, the pole $+$ continuum ansatz \cite{Hatsuda:1992bv} for the spectral densities is employed, because it adequately describes a narrow resonance, in particular the pseudo-scalar D meson.
The spectral density reads
\begin{align}\label{eq:PSansatz}
	\rho_X (s) = R_X \delta(s-m_X^2) +  \frac{1}{\pi} \mathrm{Im} \Pi^\mathrm{pert} (s) \Theta(s-s_0^X) \, ,
\end{align}
where $X$ stands for either P or S.
The residue in vacuum satisfies \cite{Narison:2003td,Bordes:2005wi,Lucha:2010ea}
\begin{align}\label{eq:fRrel}
	R_X = f_X^2 m_X^4 \, ,
\end{align}
justified by comparison of Eq.~\eqref{eq:PSansatz} with the spectral density entering the K\"{a}ll\'{e}n-Lehmann representation of the current-current correlator~\eqref{eq:ccc}, where a complete set of hadronic states has been inserted which share the quantum numbers of $X$.
In this representation, the lowest resonance enters with the residuum $|\langle 0 | j_X(0) | X \rangle|^2$ (cf.\ Eq.~\eqref{eq:deffPNarison} for the justification of~\eqref{eq:fRrel}) and further (multi-particle) excitations occur which are combined to the continuum in the ansatz~\eqref{eq:PSansatz}.
In order to render our results comparable with the findings in \cite{Narison:2003td,Bordes:2005wi,Lucha:2010ea} we assume that Eq.~\eqref{eq:fRrel} holds at finite temperatures as well.
Note that this is a simplifying assumption.
In particular, in a medium the relation~\eqref{eq:deff_Palaf_pi} might not hold anymore with the same $f_\mathrm{P}$ for $\mu=0$ and for $\mu=1,2,3$; see, e.g., \cite{Meissner:2001gz} for the corresponding case of the pion decay constant.
The reason is that a thermal medium enforces a heat-bath vector as a second four-vector in addition to the meson momentum $p^\mu$.
We will come back to a discussion of the in-medium behavior of the decay constants below.

Employing the ansatz~\eqref{eq:PSansatz} the spectral integral in Eq.~\eqref{eq:QSR} yields
\begin{align}\label{eq:PSspecInt}
	& \int\limits_0^\infty \rmd s \, e^{-s/M^2} \rho_\mathrm{X}(s) \tanh\left(\frac{\sqrt{s}}{2T}\right) \nonumber\\[-0.5ex]
	& \qquad = R_X e^{-m_X^2/M^2}\tanh\left(\frac{m_X}{2T}\right) +  \frac{1}{\pi} \int\limits_{s_0^X}^\infty \rmd s\, e^{-s/M^2} \tanh\left(\frac{\sqrt{s}}{2T}\right) \mathrm{Im} \Pi^\mathrm{pert} (s) \\[-0.0ex]
	& \qquad = \widehat{\Pi}_X^\mathrm{res} (M^2;T) + \widehat{\Pi}_X^\mathrm{cont} (M^2;T) \nonumber
\end{align}
conveniently split into the resonance part $\widehat{\Pi}_X^\mathrm{res}$ and the continuum part $\widehat{\Pi}_X^\mathrm{cont}$, being the first and second terms in the second line in~\eqref{eq:PSspecInt}, respectively.
The ratio \glspl{QSR} are utilized to deduce the meson masses from $\widetilde\Pi_X = \widehat{\Pi}_X - \widehat{\Pi}^\mathrm{cont}_X$, i.\,e.\ 
\begin{align}\label{eq:ratioQSR}
	m_X=\sqrt{\left.\left(\partial_{-M^{-2}}\widetilde\Pi_X\right)\middle/\,\widetilde\Pi_X\right.}
\end{align}
with the shorthand notation $\partial_{-M^{-2}}=-\frac{\rmd}{\rmd(1/M^2)}$ for the derivative operator, while the residua are obtained from $R_X = e^{m_X^2/M^2} \, \widetilde\Pi_X / \tanh \big( m_X/(2T) \big)$.

The part of the temperature dependence of the phenomenological side that is caused by $\tanh\big(\sqrt{s}/(2T)\big)$, cf.\ Eq.~\eqref{eq:QSR}, has only a minor numerical impact.
Even for the lowest relevant energies $\sqrt{s}$ and at temperatures $T_\mathrm{est}$ somewhat above the chiral restoration temperature the D meson masses and residua are affected on a sub-percentage level, e.\,g.\ if $T_\mathrm{est}=200\,\mathrm{MeV}$ is used for a conservative estimate:
For $\widehat{\Pi}^\mathrm{cont}_X$ entering the meson mass formula one obtains deviations of $1-\widehat{\Pi}^\mathrm{cont}_X/[\frac{1}{\pi}\int_{s_0^X}^\infty \rmd s\, e^{-s/M^2}\mathrm{Im}\Pi^\mathrm{pert}(s)] \leq 1 - \tanh \big(\sqrt{s_0^X}/(2T)\big) \leq 1 - \tanh \big(m_Q/(2T_{\mathrm{est}})\big) \simeq 0.1\,\mathrm{\%}$ for $m_Q\sim1.5\,\mathrm{GeV}$.
The numerical values of the pole residuum $R_X$ are altered by $R_X/(e^{m_X^2/M^2}\widetilde\Pi_X) - 1 = 1/\tanh\big(m_X/(2T)\big) - 1 \leq 1/\tanh\big(m_Q/(2T_{\mathrm{est}})\big) - 1 \simeq 0.1\,\mathrm{\%}$ for $m_Q\sim1.5\,\mathrm{GeV}$.
Considering the inherent uncertainties of the \gls{QSR} framework the factor $\tanh\big(\sqrt{s}/(2T)\big)$ may be safely neglected in the spectral integral kernel for our purposes, i.\,e.\ for the pole $+$ continuum ansatz~\eqref{eq:PSansatz} with physically restricted support $\{s:\sqrt{s}>m_Q\}$ at moderate temperatures $T<T_{\mathrm{est}}$.

\section{Numerical input parameters}
\label{sec:numinput}

\renewcommand{\arraystretch}{1.6}
\begin{table}[!b]
	\caption[List of condensates entering the finite temperature OPEs]{List of condensates entering the finite temperature \glspl{OPE} in Eq.~\eqref{eq:OPETdep}. The second column contains the vacuum values, while the last column provides the temperature dependent part of the condensates at $\mu=1.5\,\mathrm{GeV}$ using an \gls{RGI} chiral condensate $\hat\mu_q^3=(0.251\,\mathrm{GeV})^3$ and the integral definitions \mbox{$B_i(z) = \frac{3(3i-1)}{\pi^{2i}} \int_z^\infty \mathrm{d}y\, y^{2(i-1)}\sqrt{y^2-z^2}/(e^y-1)$} for $i=1$ and 2 \cite{Hatsuda:1992bv,Narison:2012xy}.}
	\centering
	\begin{tabular}{llr@{\hspace{0.5em}}l}
		\toprule
		condensate & vacuum value & \multicolumn{2}{l}{\hspace{0.0em}temperature dependent part}\\
		\midrule
		$\displaystyle \langle \bar q q \rangle_T$ & $\displaystyle (-0.268)^3\,\mathrm{GeV}^3$ & $\displaystyle (0.268)^3\,\mathrm{GeV}^3 \frac{T^2}{8f_\pi^2} B_1\!\!\left(\frac{m_\pi}{T}\right)\!$ & $\displaystyle \simeq\, 0.278\,\mathrm{GeV}\;T^2$	\\[1.5ex]
		$\displaystyle \langle \bar q g \sigma G q \rangle_T$ & $\displaystyle 0.8 \cdot (-0.246)^3\,\mathrm{GeV}^5$ & $\displaystyle 0.8 \cdot (0.246)^3\,\mathrm{GeV}^5 \frac{T^2}{8f_\pi^2} B_1\!\!\left(\frac{m_\pi}{T}\right)\!$ & $\displaystyle \simeq\, 0.172\,\mathrm{GeV^3}\;T^2$ \\[1.5ex]
		$\displaystyle \langle \frac{\alpha_\mathrm{s}}{\pi} G^2 \rangle_T$ & $\displaystyle 0.012\,\mathrm{GeV^4}$ &  $\displaystyle - \frac{m_\pi^2}{9} T^2 B_1\!\!\left(\frac{m_\pi}{T}\right)\!$ & $\displaystyle \simeq\, 0$ \\[1.5ex]
		$\displaystyle \langle q^\dagger i D_0 q \rangle_T$ & $\displaystyle 0$ & $\displaystyle \frac{1}{8}\!\left[ \frac{\pi^2}{5}T^4 B_2\!\!\left(\frac{m_\pi}{T}\right)\! - \frac{m_\pi^2}{8}T^2 B_1\!\!\left(\frac{m_\pi}{T}\right)\! \right] \!\!\cdot 0.916$ & $\displaystyle \simeq\, 0.247\;T^4$ \\
		\bottomrule
	\end{tabular}
	\label{tab:condTdep}
\end{table}
\renewcommand{\arraystretch}{1.0}
The numerical evaluations of the \glspl{QSR} below utilize running {QCD} parameters in the $\overline{\mathrm{MS}}$ scheme on two-loop level with $\mu = 1.5\,\mathrm{GeV}$, i.\,e.\ the strong coupling, quark mass and condensates according to Ref.~\cite{Narison:2012xy}.
The needed \gls{RGI} quantities are deduced from experimental results listed in Ref.~\cite{Olive:2016xmw}.
In particular, the {QCD} scale $\Lambda_\mathrm{QCD}$ is directly determined from $\alpha_\mathrm{s}(\mu = m_Z = 91.2\,\mathrm{GeV}) = 0.1184$ and the \gls{RGI} quark mass $\hat m_Q$ from $m_Q(\mu=2\,\mathrm{GeV})=1.275\,\mathrm{GeV}$ while the \gls{RGI} chiral condensate $\hat\mu_q^3$ employs the Gell-Mann--Oakes--Renner relation with $f_\pi=0.093\,\mathrm{GeV}$, $m_\pi=0.14\,\mathrm{GeV}$ and $(m_u+m_d)(\mu=2\,\mathrm{GeV})=0.008\,\mathrm{GeV}$.
The temperature behavior of the contributing condensates, estimated for an ambient non-interacting pion gas, has a sizable impact on the \glspl{QSR}.
The temperature dependences of numerically relevant condensates are listed in Tab.~\ref{tab:condTdep}, whereas the medium-specific condensates $\langle (\alpha_\mathrm{s}/\pi) \left[ (vG)^2/v^2 - G^2/4 \right] \rangle_T$ and $\langle \Delta \rangle_T$ \cite{Buchheim:2014eya} are negligible due to the suppression factor of order $\alpha_\mathrm{s}/\pi$ and $\text{twist}>2$, respectively \cite{Hatsuda:1992bv}.
The above decay constant $f_\pi$ and the chiral limit, i.\,e.\ $m_\pi=0$, have been used to produce the numerical coefficients in the last column in Tab.~\ref{tab:condTdep}.

\section{Conventional Borel analysis}
\label{subsec:convBA}

\subsection{Comparison of pseudo-scalar and scalar D mesons}
\label{subsubsec:compPS}

\begin{figure}[!b]
\includegraphics[trim=0mm 0mm 0mm 22mm,clip,width=0.48\textwidth]{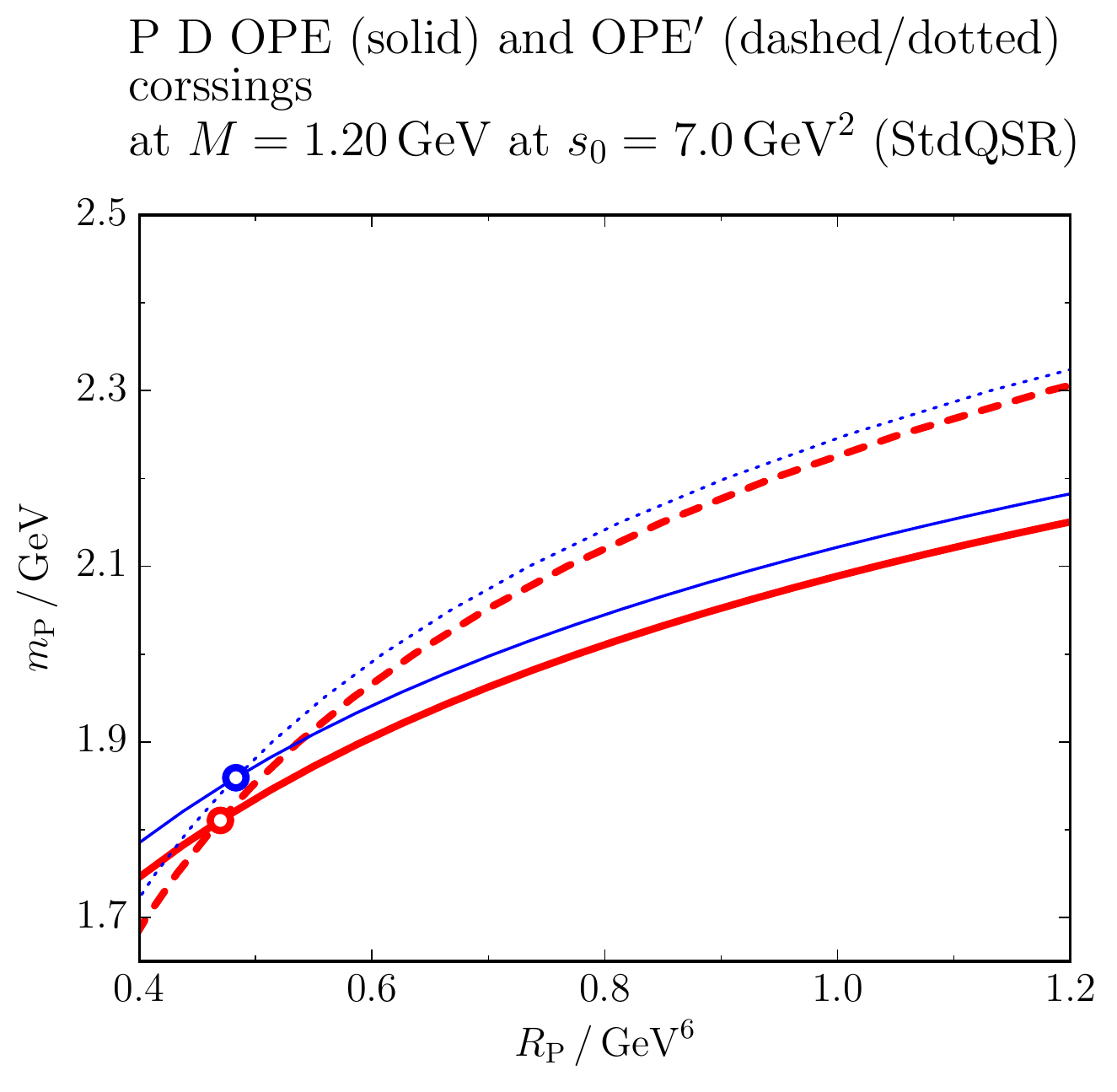}
\includegraphics[trim=0mm 0mm 0mm 22mm,clip,width=0.48\textwidth]{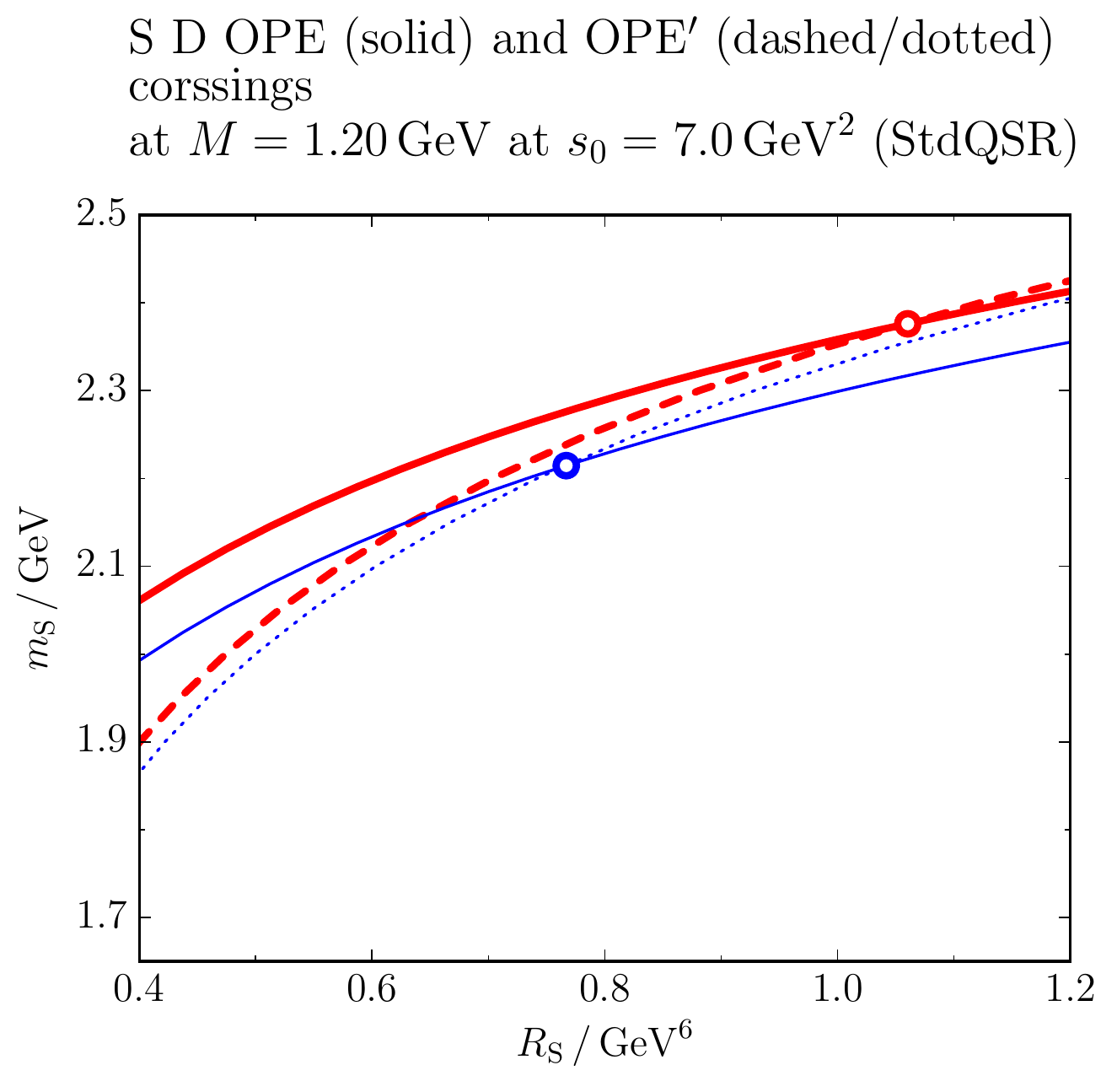}
\caption{Comparison of pseudo-scalar (left panel) and scalar (right panel) intersections of the $m_X(R_X)$ curves which originate from the sum rule $F$ (solid curves) and its derivative $F_1$ (broken curves) at $M=1.2\,\mathrm{GeV}$ and $s_0=7\,\mathrm{GeV}^2$ depicted in vacuum at $T=0$ (thick red curves) and at $T=150\,\mathrm{MeV}$ (thin blue curves).
Intersections are marked by red and blue circles, respectively.
}%
\label{fig:crossings}%
\end{figure}

Considering the Borel mass $M$ as a fiducial parameter, one has to extract the three parameters $m_X$, $R_X$ and $s_0^X$ from one equation, i.\,e.\ Eq.~\eqref{eq:QSR}, formally reading $F(m_X,R_X,s_0^X,M)=\widehat\Pi^\mathrm{res}_X(m_X,R_X,M) - \widetilde\Pi_X(s_0^X,M) = 0$.
There are several strategies to accomplish that goal, cf.\ Refs.~\cite{Reinders:1984sr,Leinweber:1995fn}.
We follow here as closely as possible our previously employed procedure for the $\uprho$ meson \cite{Zschocke:2002mn}, i.\,e.\ we supplement Eq.~\eqref{eq:QSR} by the derivative sum rule emerging from $\partial_{-M^{-2}}$\eqref{eq:QSR}, formally $\partial_{-M^{-2}}F = F_1(m_X,R_X,s_0^X,M)=0$.
(This is a common procedure, see \cite{Shifman:1978bx,Reinders:1984sr} for reasoning and applications.)
The resulting two equations can be combined to obtain $m_X(s_0^X,M)$ and $R_X(s_0^X,M)$.
Prior to that, the curves $m_X(R_X;s_0^X,M)$ have to be constructed from $F(m_X,R_X,s_0^X,M)=0$ and $F_1(m_X,R_X,s_0^X,M)=0$ at given values of $s_0^X$ and $M$.
Examples are exhibited in Fig.~\ref{fig:crossings} for both the pseudo-scalar (left) and scalar (right) channels as well as for $T=0$ and $T=150\,\mathrm{MeV}$.
The displayed curves give some impression of the temperature impact (compare red and blue curves) on the $F$ and $F_1$ sum rules (compare solid and broken curves).
Note the opposite ordering of the corresponding curves for pseudo-scalar and scalar channels.

\begin{figure}[!t]
\settoheight{\imageheight}{\includegraphics[trim=0mm 0mm 0mm 15mm,clip,width=0.48\textwidth]{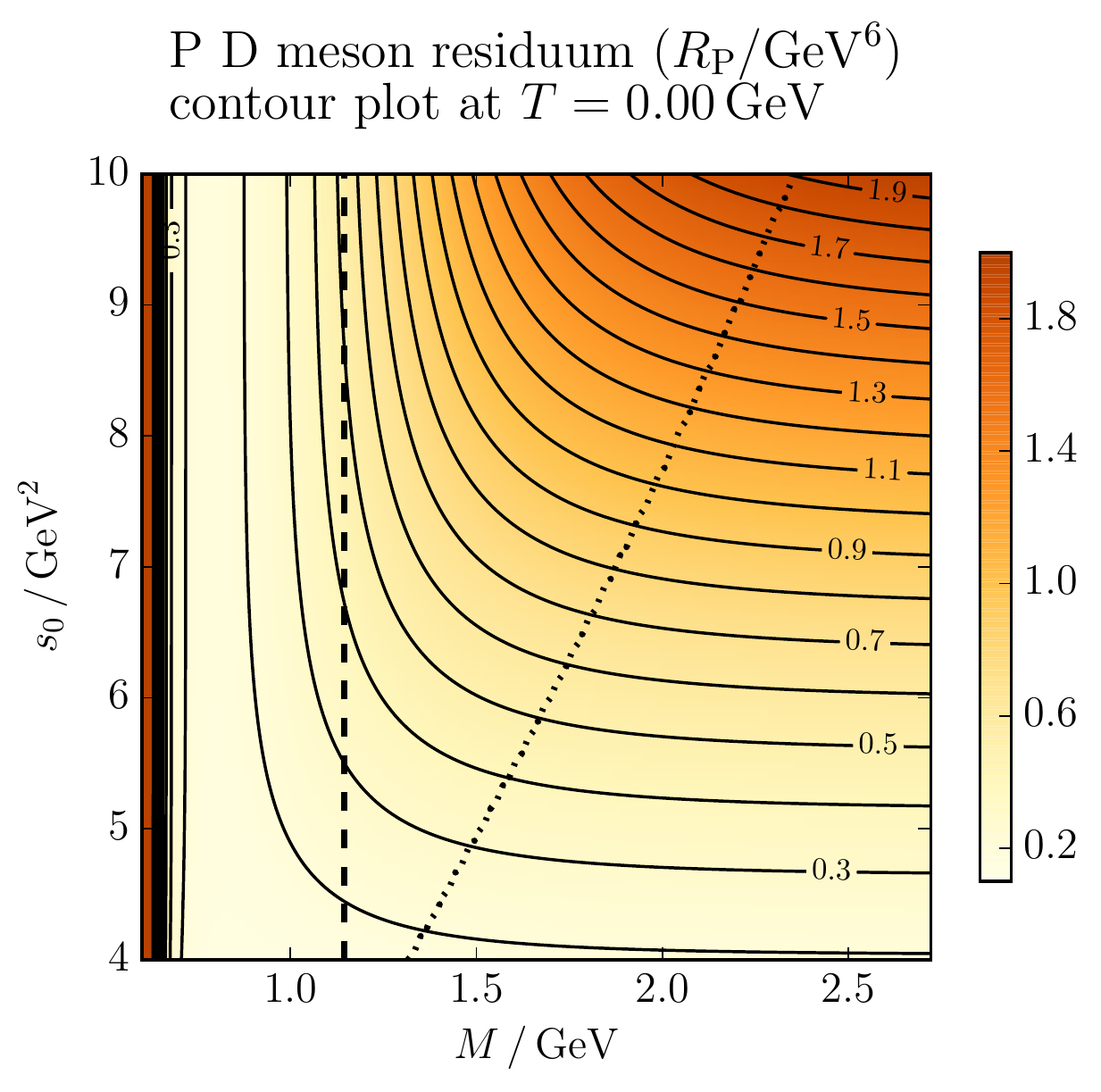}}
\centering
\begin{minipage}{0.48\textwidth}
\begin{flushleft}
	\includegraphics[trim=0mm 0mm 15mm 15mm,clip,height=\imageheight]{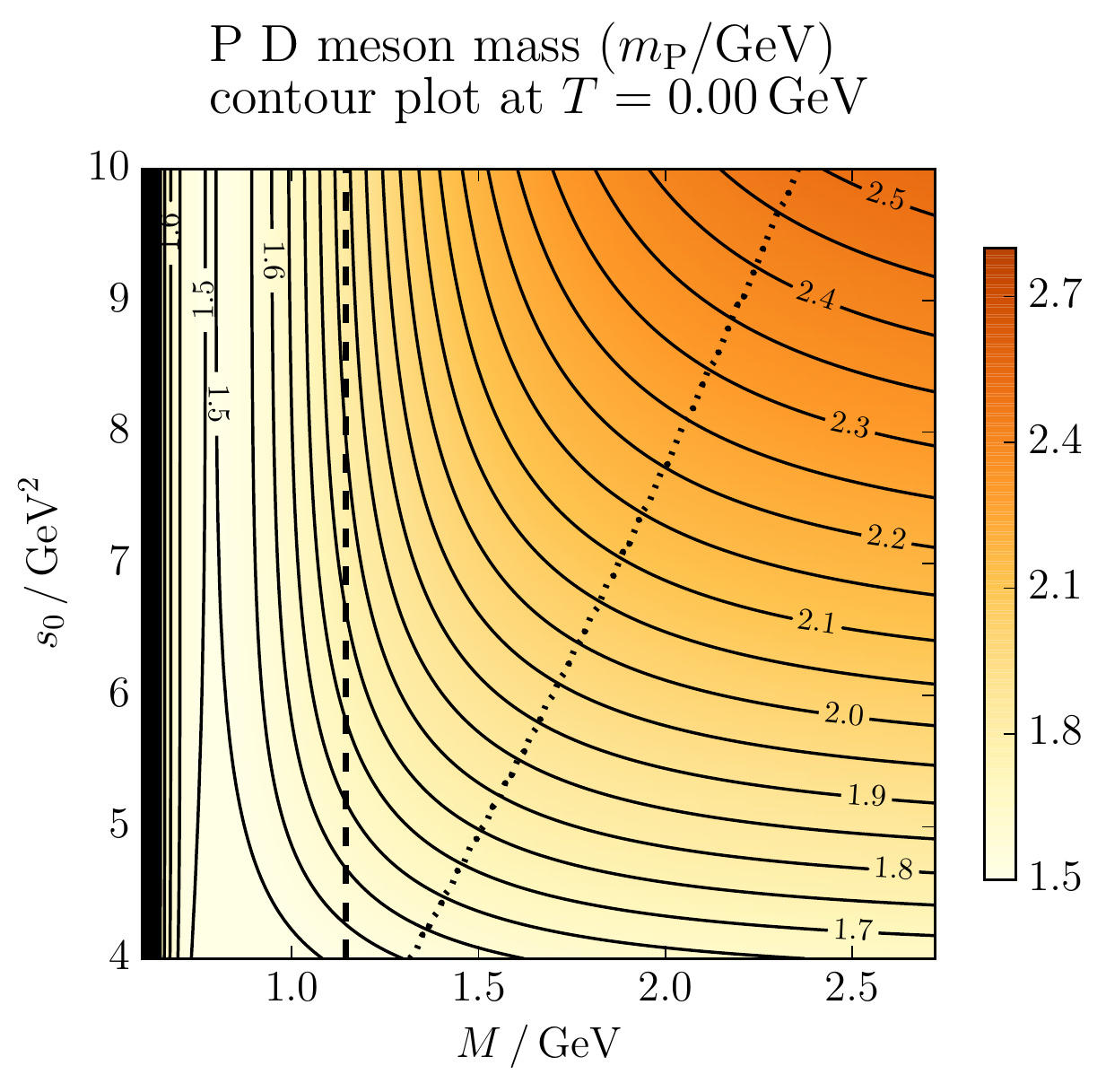}
	\includegraphics[trim=0mm 0mm 15mm 15mm,clip,height=\imageheight]{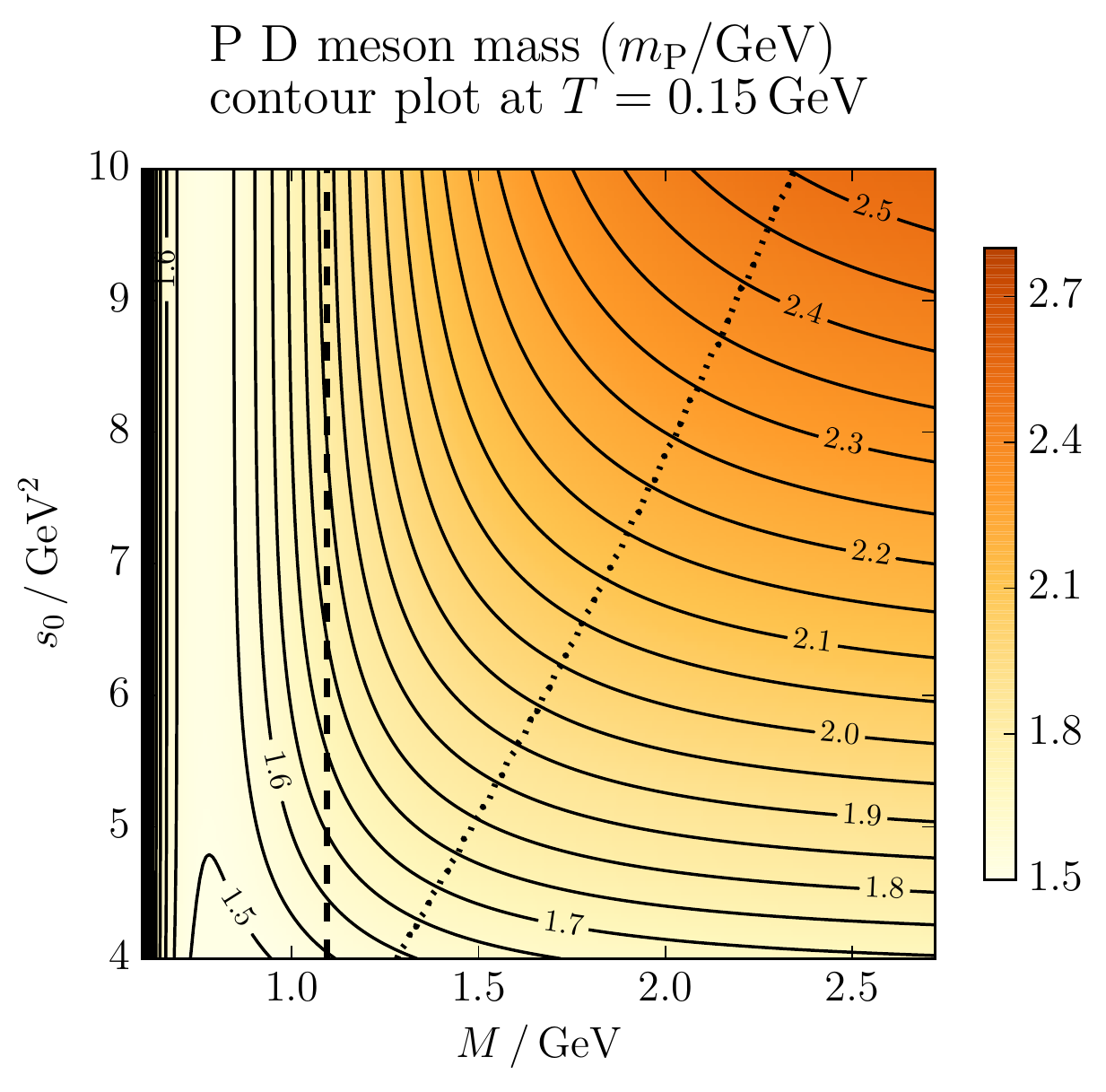}
\end{flushleft}
\end{minipage}
\hspace{-3.0em}
\begin{minipage}{0.48\textwidth}
\begin{flushleft}
	\includegraphics[trim=0mm 0mm 15mm 15mm,clip,height=\imageheight]{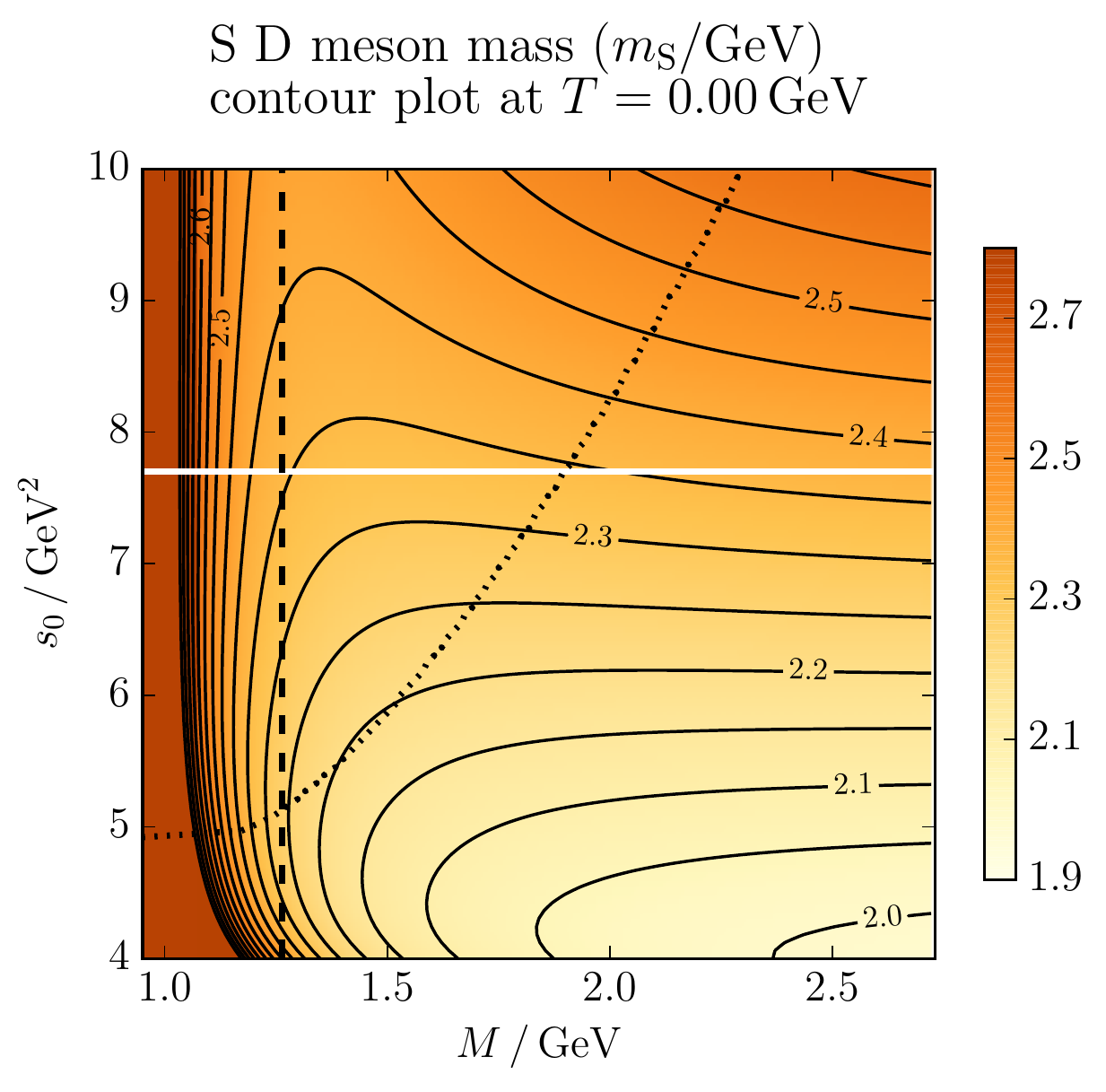}
	\includegraphics[trim=0mm 0mm 15mm 15mm,clip,height=\imageheight]{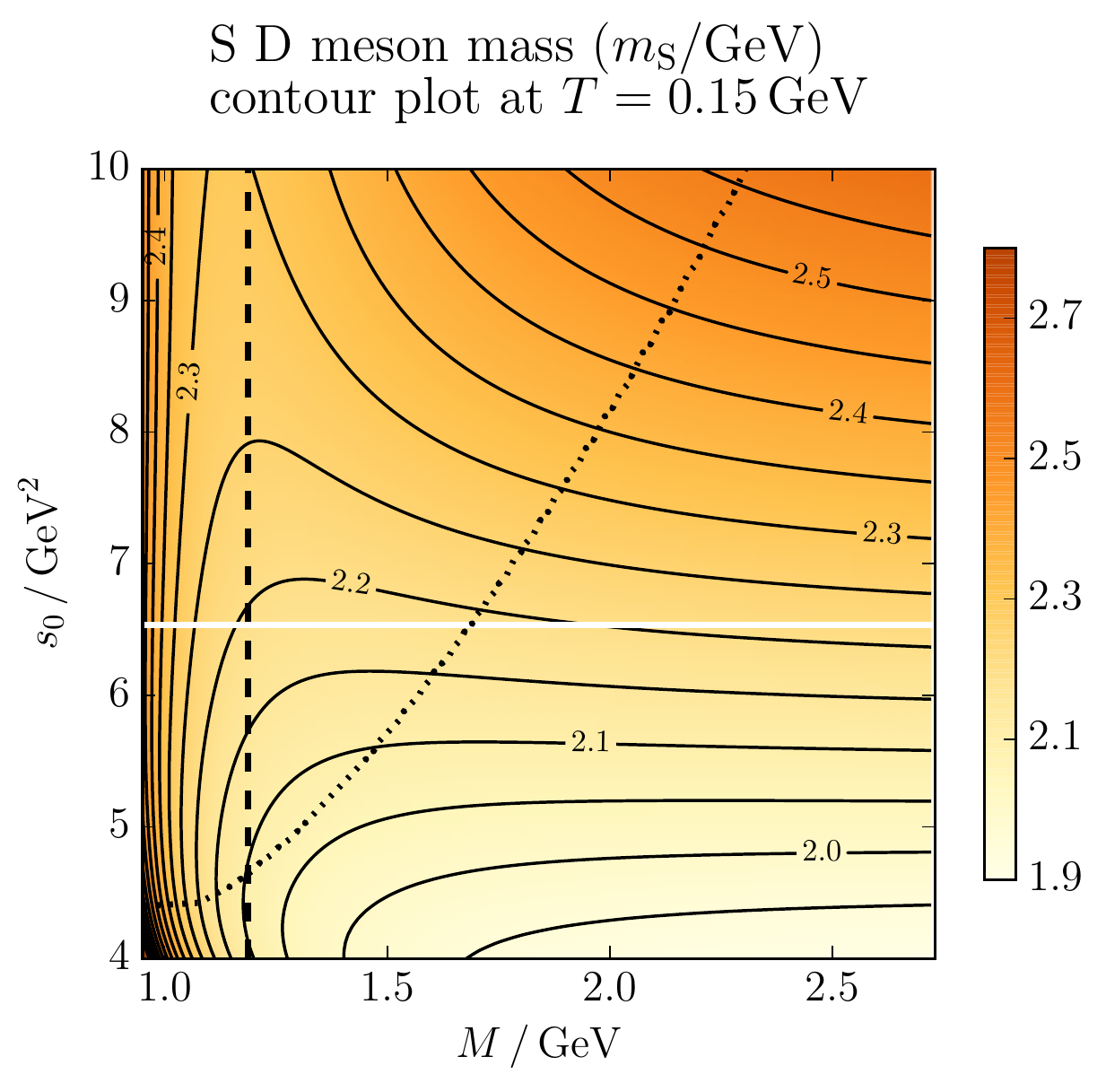}
\end{flushleft}
\end{minipage}
\caption[Comparison of pseudo-scalar and scalar D meson mass contours]{Comparison of pseudo-scalar (left panels) and scalar (right panels) D meson mass contours $m_\mathrm{P,S}(s_0,M)\,/\,\mathrm{GeV}$, in vacuum at $T=0$ (upper panels) and at $T=150\,\mathrm{MeV}$ (lower panels).
The dashed and dotted curves depict the lower and upper Borel window boundaries, respectively.
The white bars in the right panels mark the continuum threshold parameters resulting from the numerical analysis in Sec.~\ref{subsubsec:numS}.}%
\label{fig:PSm(M,s)StdBwin}%
\end{figure}
The intersections (circles in Fig.~\ref{fig:crossings}) of the $F$ and $F_1$ curves, at one temperature, deliver the respective values of $m_X$ and $R_X$. Scanning over the $s_0$-$M$ plane yields the wanted surfaces $m_X(s_0,M)$ and $R_X(s_0,M)$.
These quantities are exhibited in Figs.~\ref{fig:PSm(M,s)StdBwin} and \ref{fig:PSR(M,s)StdBwin} as contour plots.
The upper and lower limits of the Borel windows, $M^X_\mathrm{max,min}(s_0)$, are also shown.
This window is the Borel mass range where the phenomenological and \gls{OPE} sides of the \gls{QSR} can be reliably matched (to some extent), since it is constructed such that higher \gls{OPE} terms do not contribute significantly and the phenomenological spectral density is dominated by the first excitation \cite{Leinweber:1995fn}.
In the \gls{QSR} framework, meson masses are evaluated as the average of the particular meson mass Borel curve $m_X(M)$ which shows maximum flatness in the corresponding Borel window.

\begin{figure}[t]
\settoheight{\imageheight}{\includegraphics[trim=0mm 0mm 0mm 15mm,clip,width=0.48\textwidth]{fig3a.pdf}}
\centering
\begin{minipage}{0.48\textwidth}
\begin{flushleft}
	\includegraphics[trim=0mm 0mm 15mm 15mm,clip,height=\imageheight]{fig3a.pdf}
	\includegraphics[trim=0mm 0mm 15mm 15mm,clip,height=\imageheight]{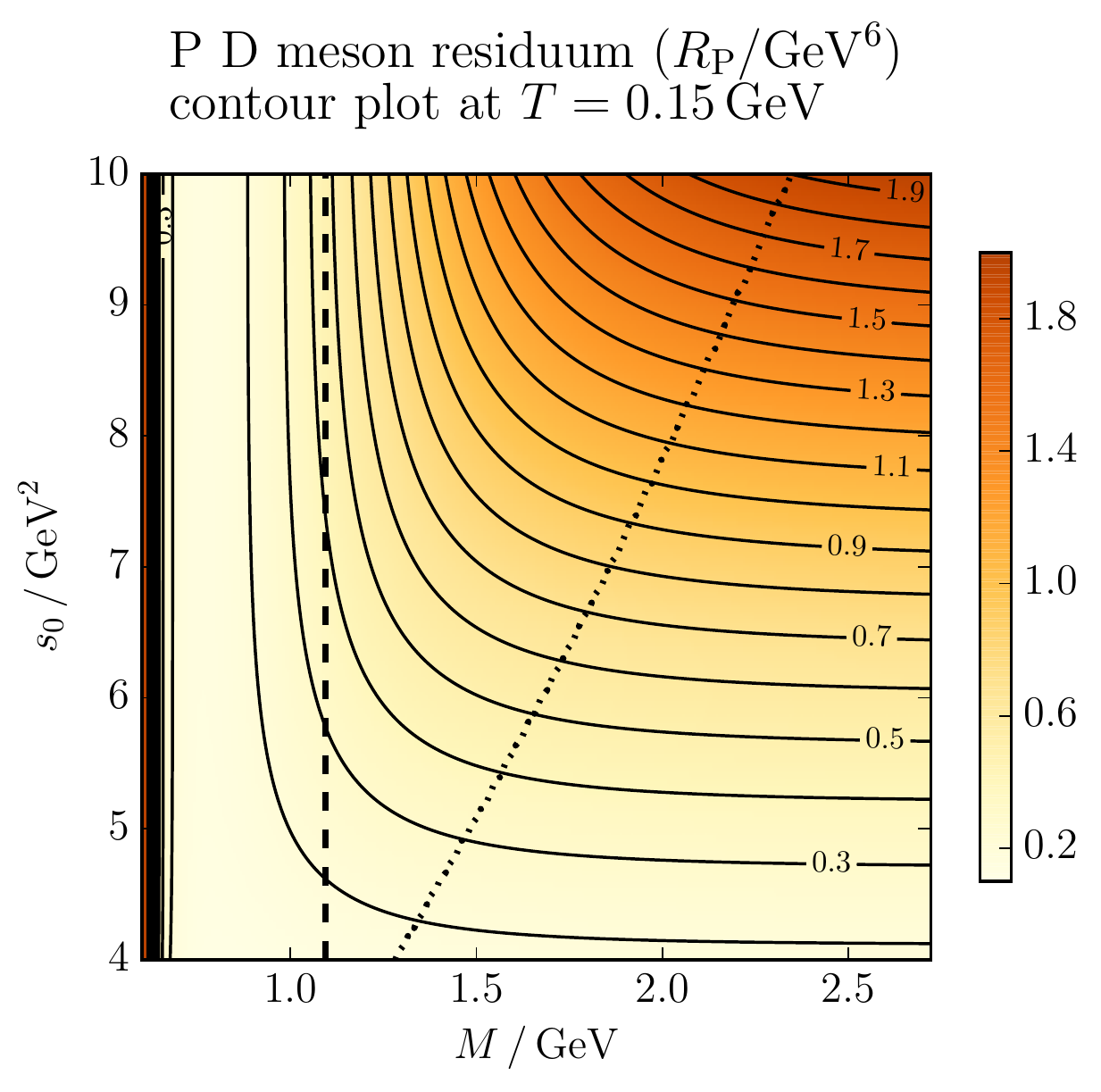}
\end{flushleft}
\end{minipage}
\hspace{-3.0em}
\begin{minipage}{0.48\textwidth}
\begin{flushleft}
	\includegraphics[trim=0mm 0mm 15mm 15mm,clip,height=\imageheight]{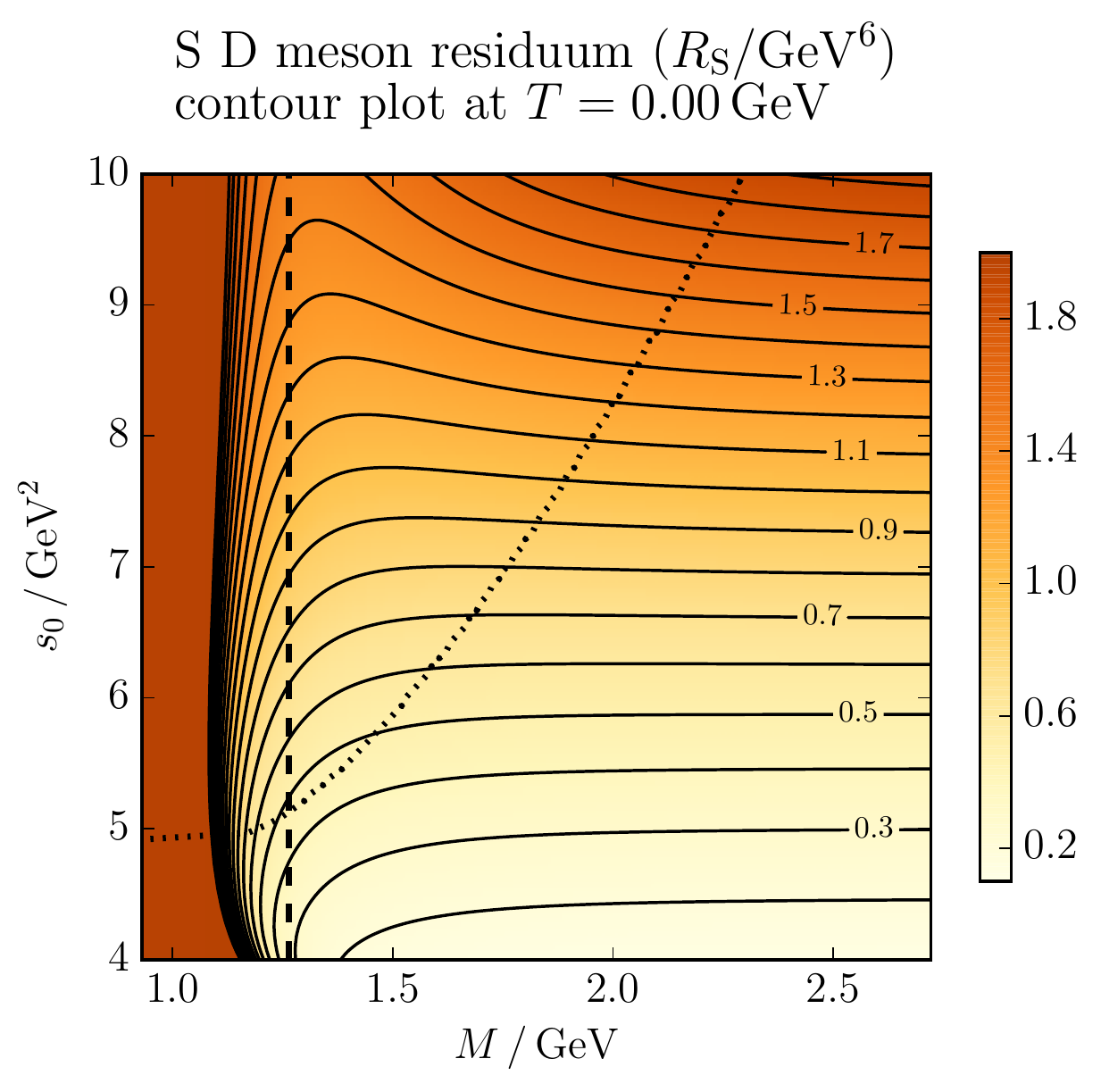}
	\includegraphics[trim=0mm 0mm 15mm 15mm,clip,height=\imageheight]{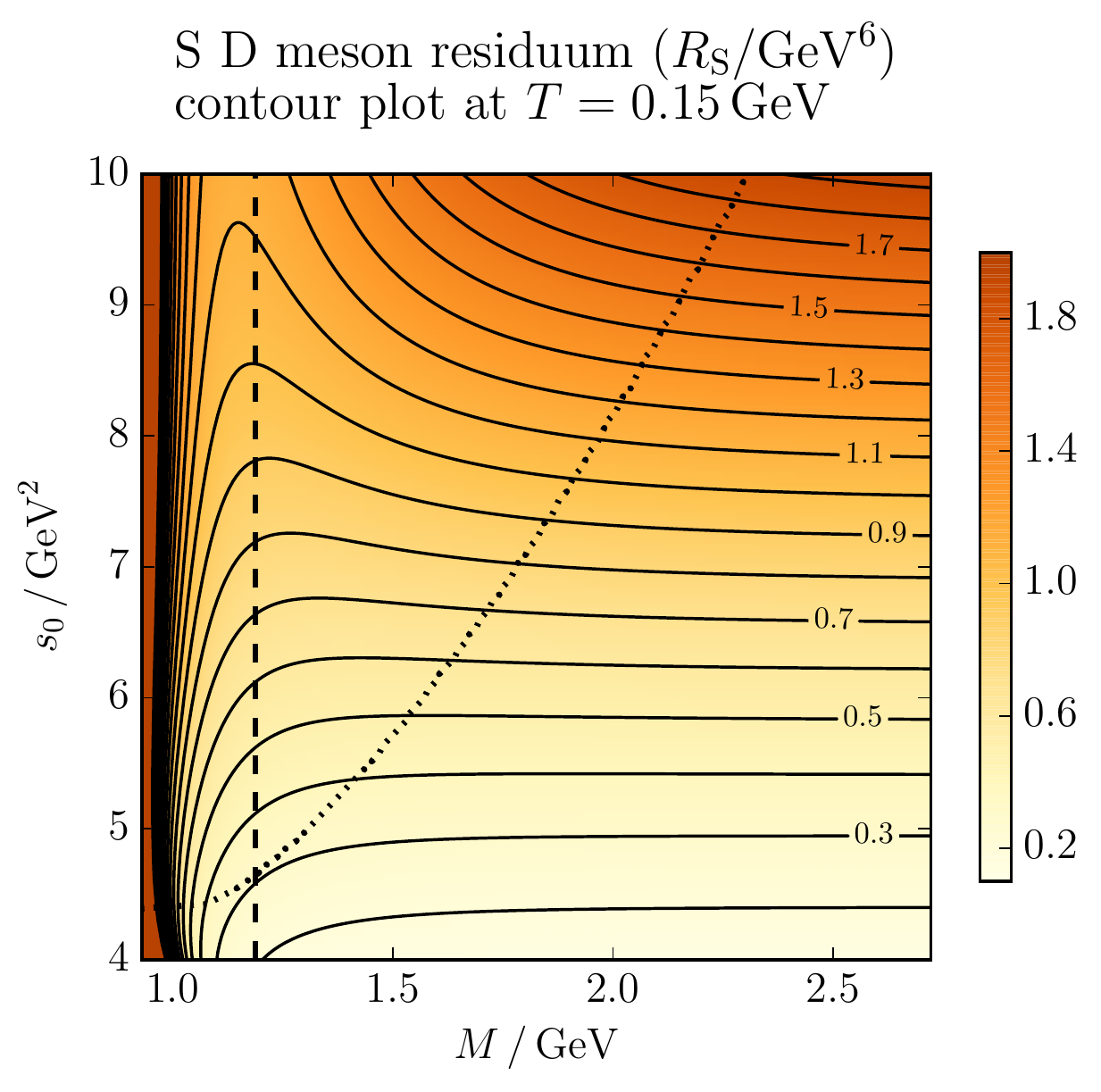}
\end{flushleft}
\end{minipage}
\caption[Comparison of pseudo-scalar and scalar D meson residuum contours]{Comparison of pseudo-scalar (left panels) and scalar (right panels) D meson residuum contours $R_\mathrm{P,S}(s_0,M)\,/\,\mathrm{GeV}^6$, in vacuum at $T=0$ (upper panels) and at $T=150\,\mathrm{MeV}$ (lower panels). The dashed and dotted curves depict the lower and upper Borel window boundaries, respectively.}%
\label{fig:PSR(M,s)StdBwin}%
\end{figure}
The left panels in Figs.~\ref{fig:PSm(M,s)StdBwin} and \ref{fig:PSR(M,s)StdBwin} unravel an unpleasant feature for the pseudo-scalar D meson: It seems hardly possible to identify a horizontally flat section of $m_\mathrm{P}(s_0,M)$ within the Borel window which allows pinning down a robust value of the threshold $s_0^\mathrm{P}$.
Going to lower values of $s_0^\mathrm{P}$ would $(i)$ violate the requirement $m_\mathrm{P}^2 < s_0^\mathrm{P}$ and $(ii)$ run into danger of a closing Borel window.
The large-$M$ region is either beyond the Borel window or/and already in the insensitive
perturbative region.
We argue that due to this reasoning the pseudo-scalar D meson Borel analysis is hampered by such peculiarities,%
\footnote{We have tested that evaluations combining weighted finite energy sum rules, cf.~\cite{Maltman:1999rh,Steinmueller:2006id}, and the genuine Borel sum rule \eqref{eq:QSR} to fix the spectral parameters do not eliminate such peculiarities.}
 which have been circumvented in \cite{Hilger:2008jg,Narison:2012xy} by some special handling.%
\footnote{In Ref.~\cite{Hilger:2008jg} the threshold parameter $s_0^\mathrm{P}$ is chosen by hand to produce the experimental vacuum D meson mass, where a linear density dependence of $s_0^\mathrm{P}$ is used to monitor the influence of this parameter on the D meson mass at finite density.
In Ref.~\cite{Narison:2012xy} the Borel curves of the \gls{QSR} analysis are stabilized by a floating renormalization scale $\mu = M$.}
The situation for the scalar D meson (right panels in Fig.~\ref{fig:PSm(M,s)StdBwin}) is much more suitable for adjusting the threshold and extracting mass and residuum.
This analysis will be carried out in Sec.~\ref{subsubsec:numS}.
The differences of the temperature effects on pseudo-scalar and scalar channels are to be discussed in Sec.~\ref{subsubsec:originTeff}.

We emphasize that for $T=0$ (upper panels in Figs.~\ref{fig:PSm(M,s)StdBwin} and \ref{fig:PSR(M,s)StdBwin}) and $T=150\,\mathrm{MeV}$ (lower panels in Figs.~\ref{fig:PSm(M,s)StdBwin} and \ref{fig:PSR(M,s)StdBwin}) the overall situation is the same.
However, while in the pseudo-scalar channel the temperature effects within the region of interest are small (cf.\ left panels in Fig.~\ref{fig:PSmR(M,s)dT}), in the scalar channel the impact of non-zero temperatures is more pronounced in particular towards lower values of $M$ (cf.\ right panels in Fig.~\ref{fig:PSmR(M,s)dT}), where the mass parameter $m_\mathrm{S}$ drops by about $150\,\mathrm{MeV}$, and the residue $R_\mathrm{S}$ by $0.2\,\mathrm{GeV^6}$.
This is to be contrasted with $m_\mathrm{P}$ slightly growing by $m_\mathrm{P}=50\,\mathrm{MeV}$ at most, and a nearly unaffected residuum, i.\,e.\ $R_\mathrm{P}|_{T=0}\approx R_\mathrm{P}|_{T=150\,\mathrm{MeV}}$, which can be read off from Fig.~\ref{fig:PSmR(M,s)dT}.
However, we cannot pin down reliable values of $m_\mathrm{P}$ and subsequently $R_\mathrm{P}$ within the conventional approach.
To summarize, the scenario $m_\mathrm{S} \rightarrow m_\mathrm{P}$ is the qualitative outcome of our study using the conventional \gls{QSR} approach.
This outcome is in agreement with the findings in \cite{Sasaki:2014asa}.
However, we will find quantitative differences in Sec.~\ref{subsec:LuchaMedS}.
\begin{figure}[!t]
\centering
\includegraphics[trim=0mm 0mm 0mm 15mm,clip,width=0.48\textwidth]{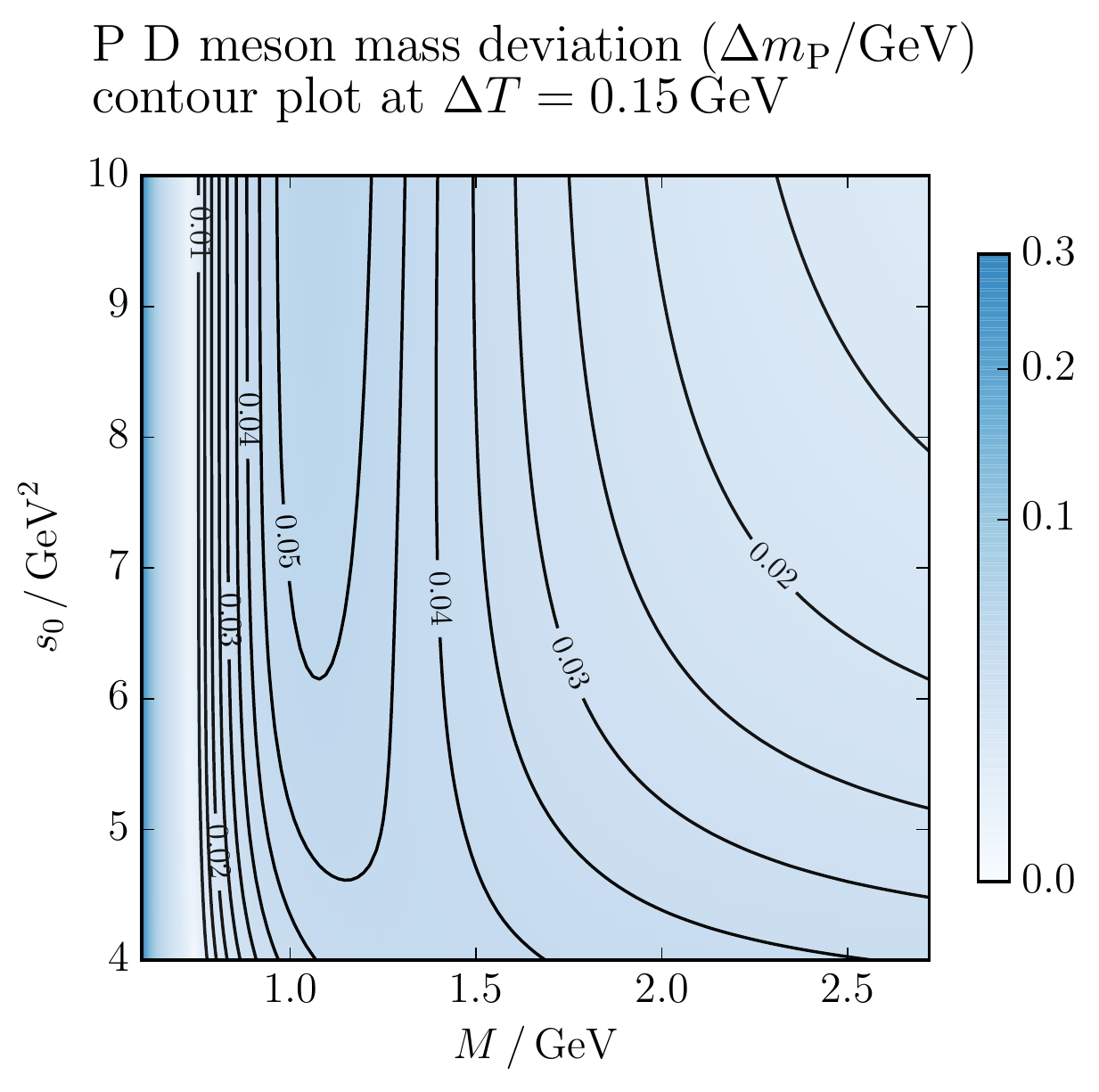}
\hspace{-3.5em}
\includegraphics[trim=0mm 0mm 0mm 15mm,clip,width=0.48\textwidth]{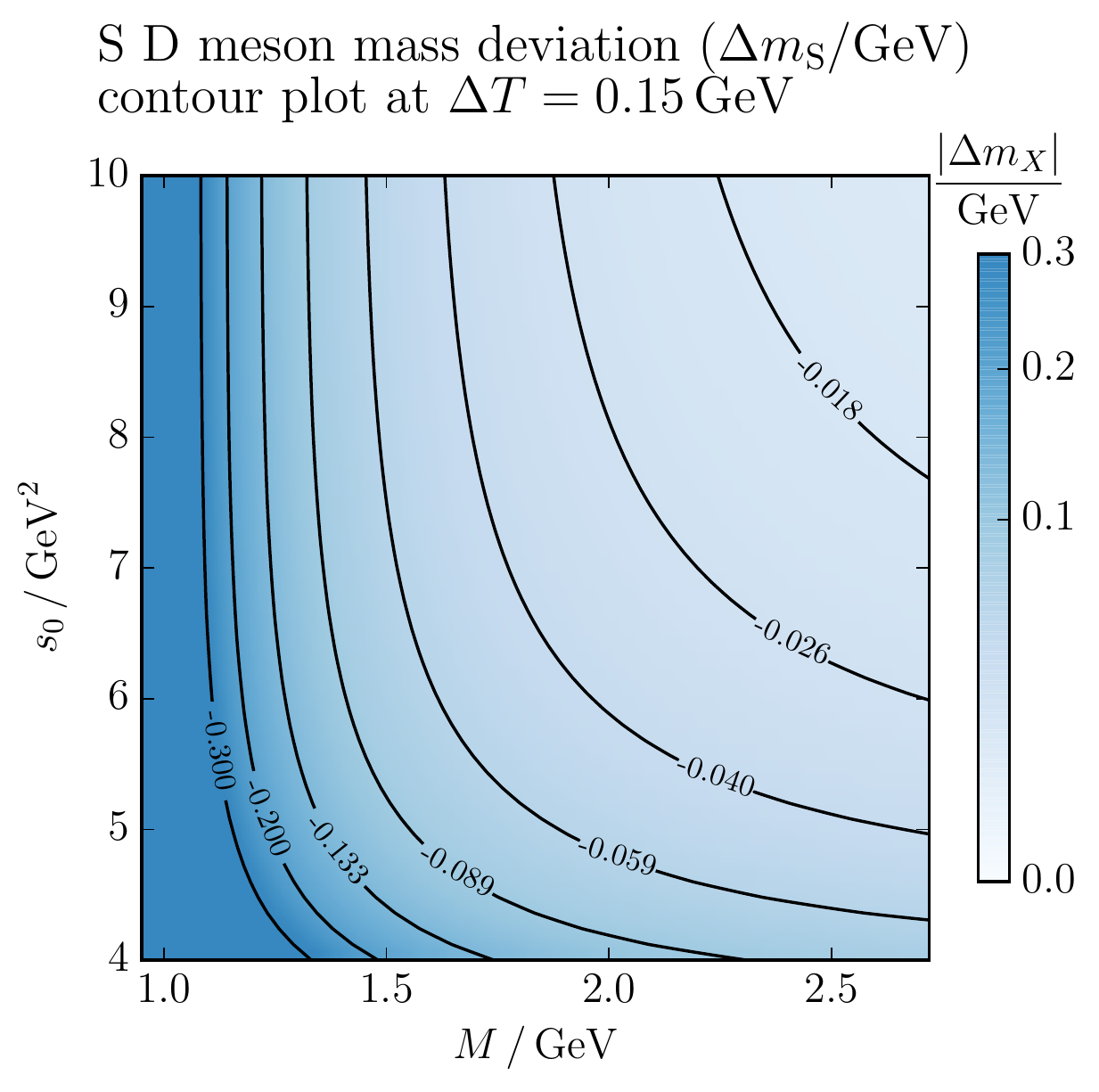}
\includegraphics[trim=0mm 0mm 0mm 15mm,clip,width=0.48\textwidth]{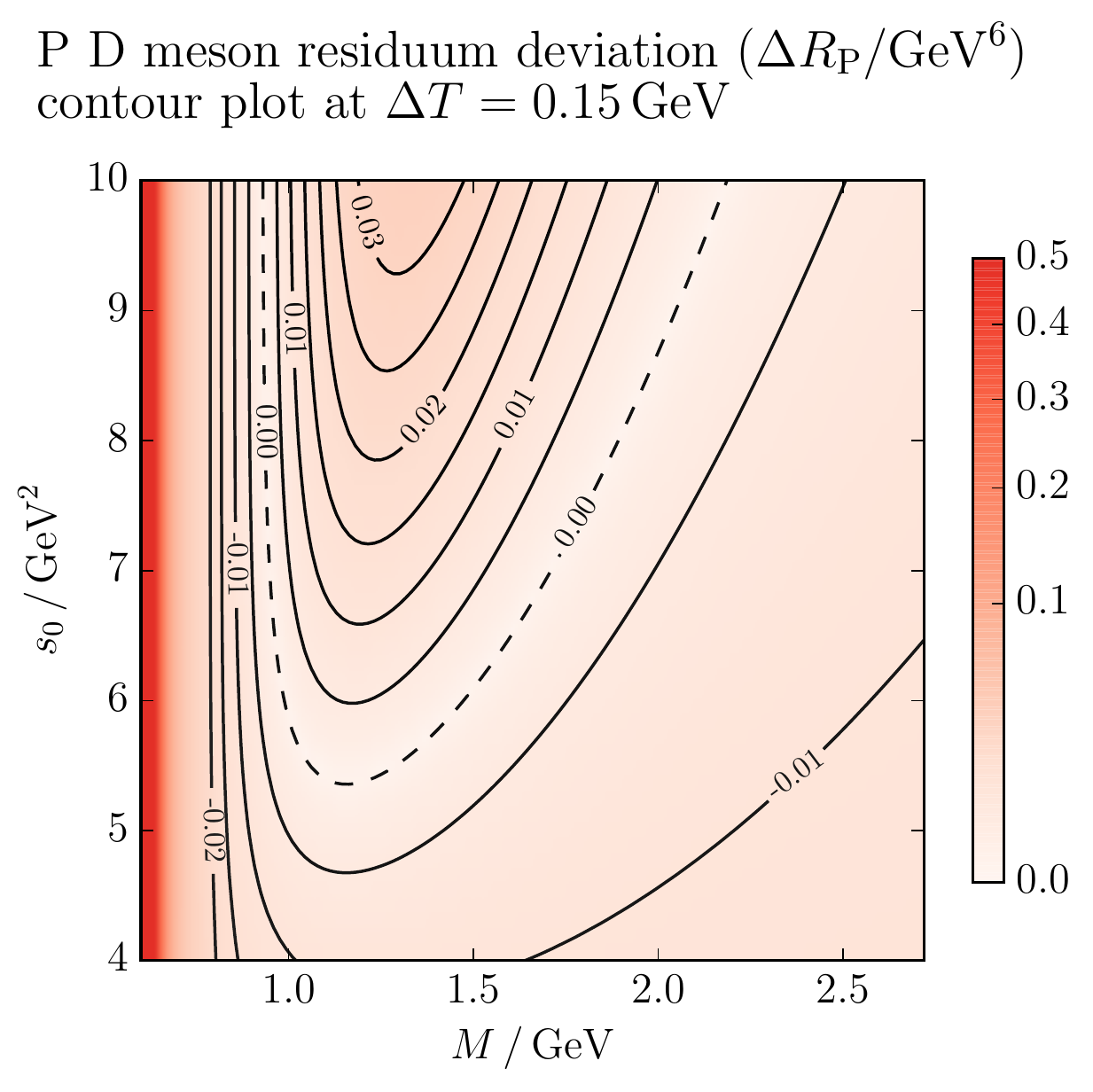}
\hspace{-3.5em}
\includegraphics[trim=0mm 0mm 0mm 15mm,clip,width=0.48\textwidth]{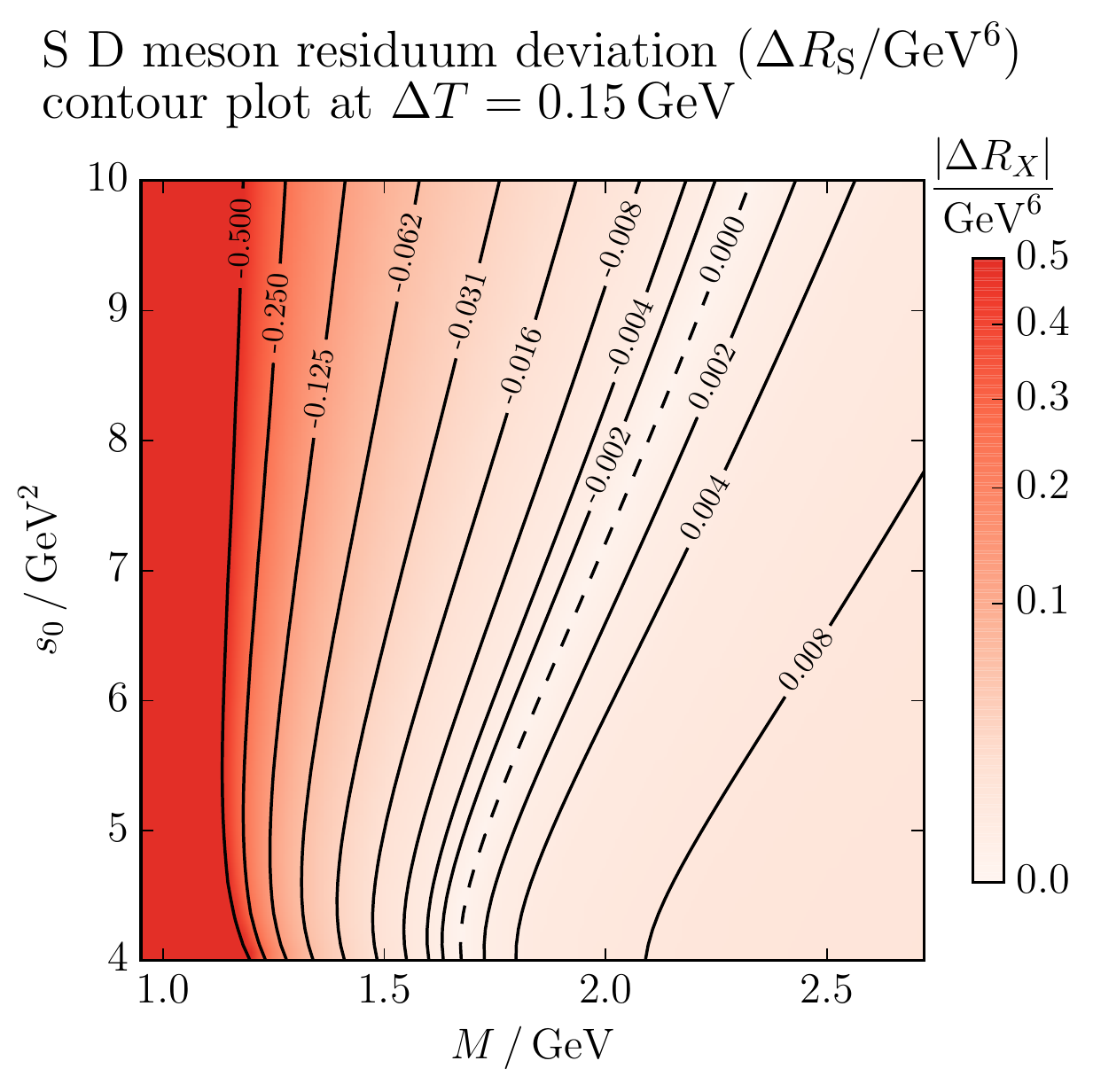}
\caption[Comparison of pseudo-scalar and scalar D meson mass difference and residuum difference contours]{Comparison of pseudo-scalar (left panels) and scalar (right panels) D meson mass difference contours $\Delta m_\mathrm{P,S}(s_0,M)\,/\,\mathrm{GeV}$ ($\Delta m_X(s_0,M) = m_X(s_0,M)|_{T=150\,\mathrm{MeV}} - m_X(s_0,M)|_{T=0}$, upper panels) and residuum difference contours $\Delta R_\mathrm{P,S}(s_0,M)\,/\,\mathrm{GeV}^6$ ($\Delta R_X(s_0,M) = R_X(s_0,M)|_{T=150\,\mathrm{MeV}} - R_X(s_0,M)|_{T=0}$, lower panels), respectively. 
Note that the increment between contour lines in the right panels is not constant in contrast to the left panels and any other contour plot in this paper.
In order to elucidate the numerical impact of the temperature differences, the contour colors depict the modulus of the differences while the contour lines are labeled with the corresponding sign.
}%
\label{fig:PSmR(M,s)dT}%
\end{figure}

\subsection{Scalar D meson}
\label{subsubsec:numS}

\begin{figure}[!t]
\settoheight{\imageheight}{\includegraphics[trim=0mm 0mm 0mm 15mm,clip,width=0.48\textwidth]{fig3a.pdf}}
\centering
\begin{minipage}{0.48\textwidth}
\begin{flushleft}
\hspace{2em}
\includegraphics[trim=0mm 0mm 0mm 14mm,clip,height=\imageheight]{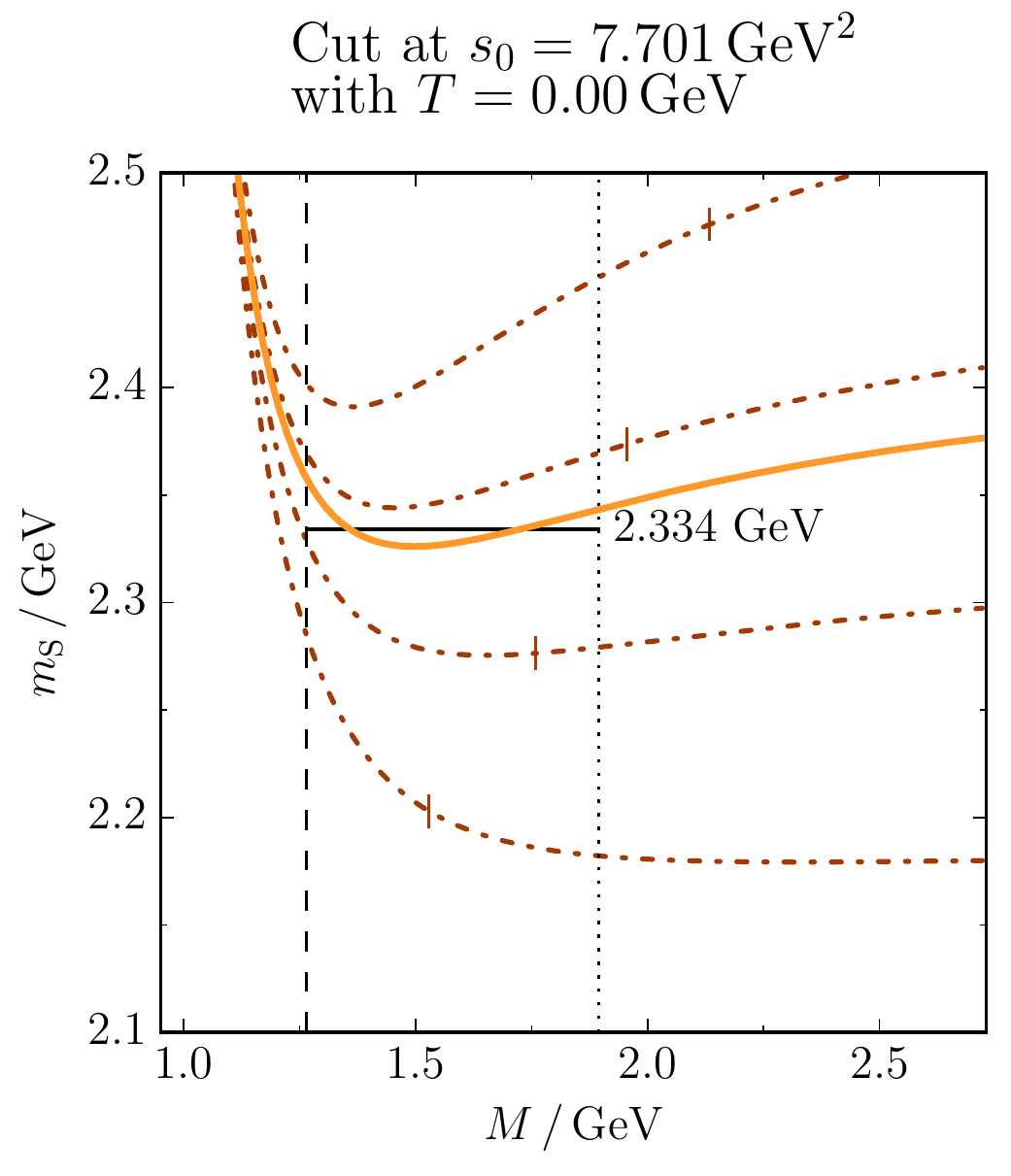}
\end{flushleft}
\end{minipage}
\hspace{1mm}
\begin{minipage}{0.48\textwidth}
\begin{flushleft}
\includegraphics[trim=0mm 0mm 0mm 14mm,clip,height=\imageheight]{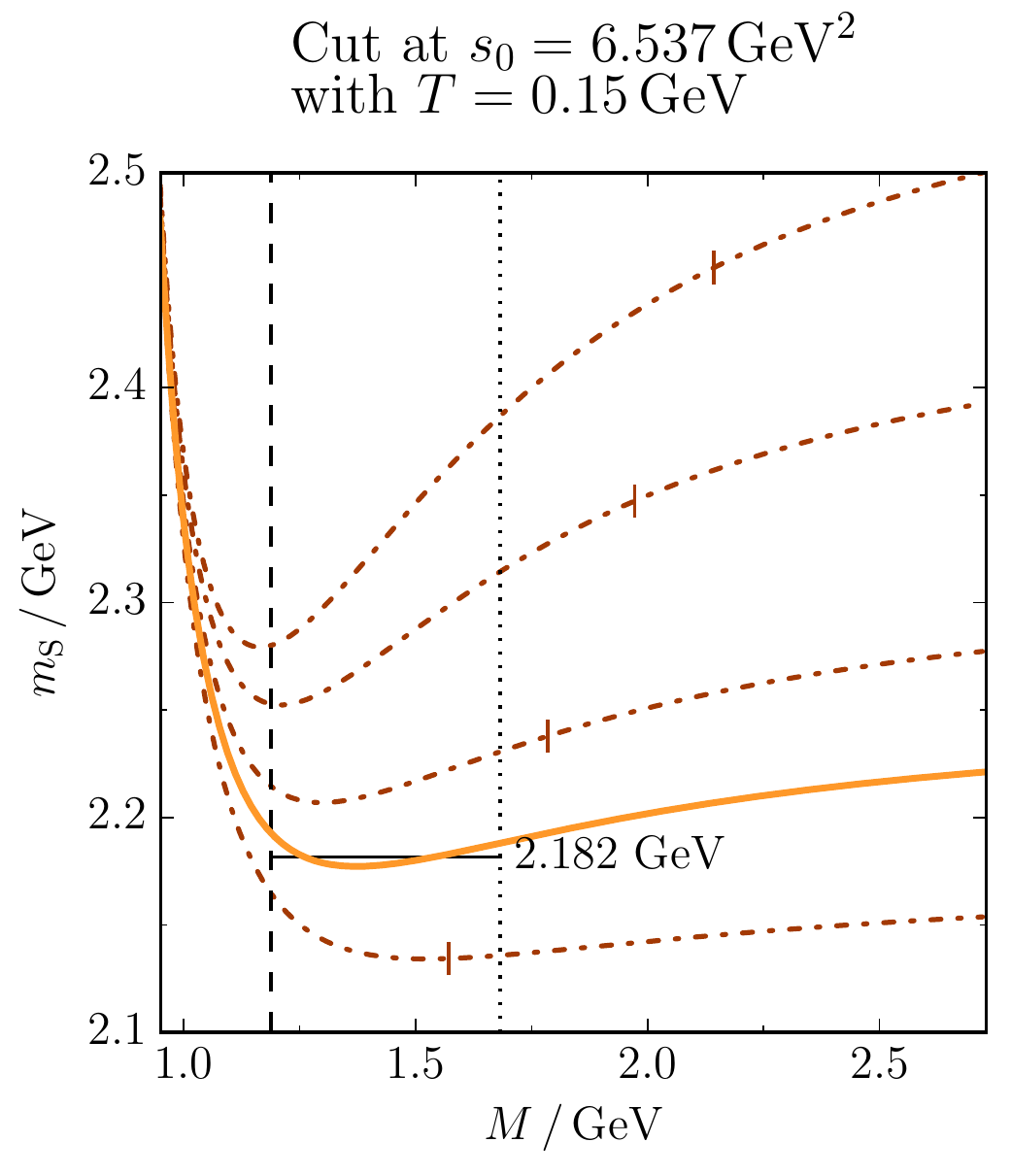}
\end{flushleft}
\end{minipage}
\caption[Optimized scalar D meson mass mass Borel curves]{The optimized scalar D meson mass Borel curves, in vacuum at $T=0$ (left panel) and at $T=150\,\mathrm{MeV}$ (right panel). These Borel curves (orange solid) are those curves $m_\mathrm{S}$ as a function of $M$ which are maximally flat within the respective Borel window by selecting an appropriate value of $s_0^\mathrm{S}$, i.\,e.\ they represent cross sections through the landscapes in the right panels of Fig.~\ref{fig:PSm(M,s)StdBwin} along the white horizontal lines.
The mean values are depicted by black horizontal lines.
The optimum values are $s_0^\mathrm{S} = 7.69$ $(6.54)\,\mathrm{GeV^2}$ for $T=0$ and $T=150\,\mathrm{MeV}$, respectively.
For comparison, the mass Borel curves (brown dot-dashed) at $s_0^\mathrm{S}=6$, 7, 8, and $9\,\mathrm{GeV}^2$ are displayed (from bottom to top), where the short vertical lines mark the corresponding $M_\mathrm{max}^\mathrm{S}(s_0^\mathrm{S})$.
The steep rise of the Borel curves at small values of $M$ point to some uncomfortable sensitivity against variations of the lower Borel window limit.
}%
\label{fig:Sm(M,s)cuts}%
\end{figure}
A further quantitative analysis, which is based on the conventional approach requiring maximal flatness of the mass Borel curve within the Borel window, is worthwhile.
The steep Borel curve section spoiling the extraction method of the spectral meson parameters as well as the insensitive perturbative region are avoided if standard criteria are imposed upon the Borel window range \cite{Leinweber:1995fn}.
Therefore, the lower limit $M^\mathrm{S}_\mathrm{min}$ of the Borel window, where the highest order condensate term is required to contribute less than $10\,\mathrm{\%}$ to the \gls{OPE}, is determined from the sum of the moduli of the dimension-5 terms; and the upper limit $M^\mathrm{S}_\mathrm{max}$ is extracted by requiring the continuum to contribute less than $50\,\mathrm{\%}$ to the spectral integral of the \gls{QSR}.

The results exhibit a mass drop of the scalar D meson.
As depicted in Fig.~\ref{fig:Sm(M,s)cuts}, calculating the average of the mass Borel curve with optimal continuum threshold parameter, corresponding to the white cuts in the right panels in Fig.~\ref{fig:PSm(M,s)StdBwin}, yields the scalar D meson mass $m_\mathrm{S}=2.334\,\mathrm{GeV}$ (which compares well with experimental value $m_\mathrm{S}=2.318\,\mathrm{GeV}$ \cite{Olive:2016xmw}) in vacuum which drops to $m_\mathrm{S}=2.182\,\mathrm{GeV}$ at $T=150\,\mathrm{MeV}$.
This mass drop is caused essentially by the changed continuum threshold parameter which in turn comes from the mildly changed contours of $m_\mathrm{S}(s_0,M)$ due to temperatures effects to be discussed further below in Sec.~\ref{subsubsec:originTeff}.

\begin{figure}[!t]
\settoheight{\imageheight}{\includegraphics[trim=0mm 0mm 0mm 17mm,clip,width=0.49\textwidth]{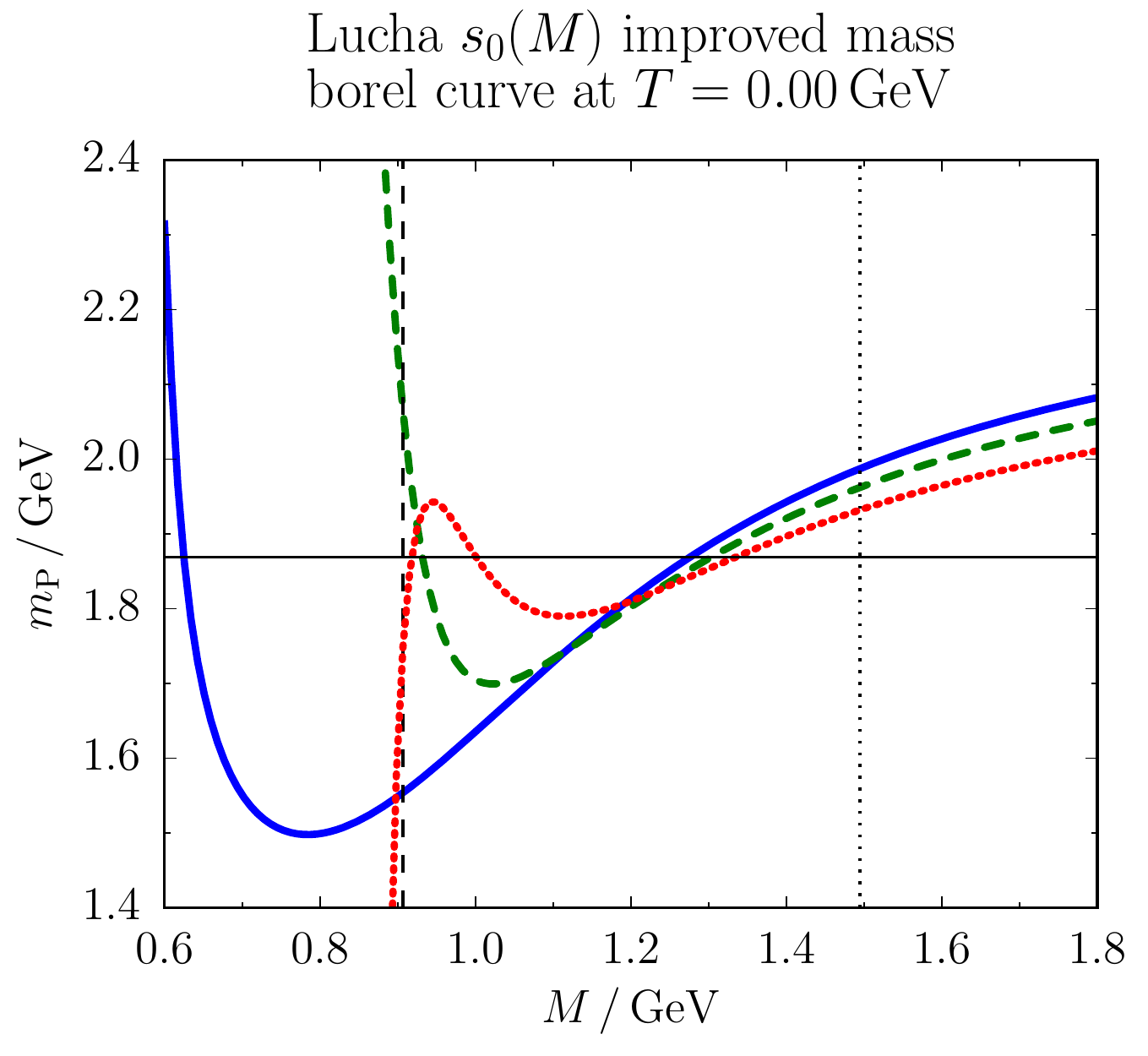}}%
\includegraphics[trim=1.5mm 0mm 1.2mm 10mm,clip,height=0.98\imageheight]{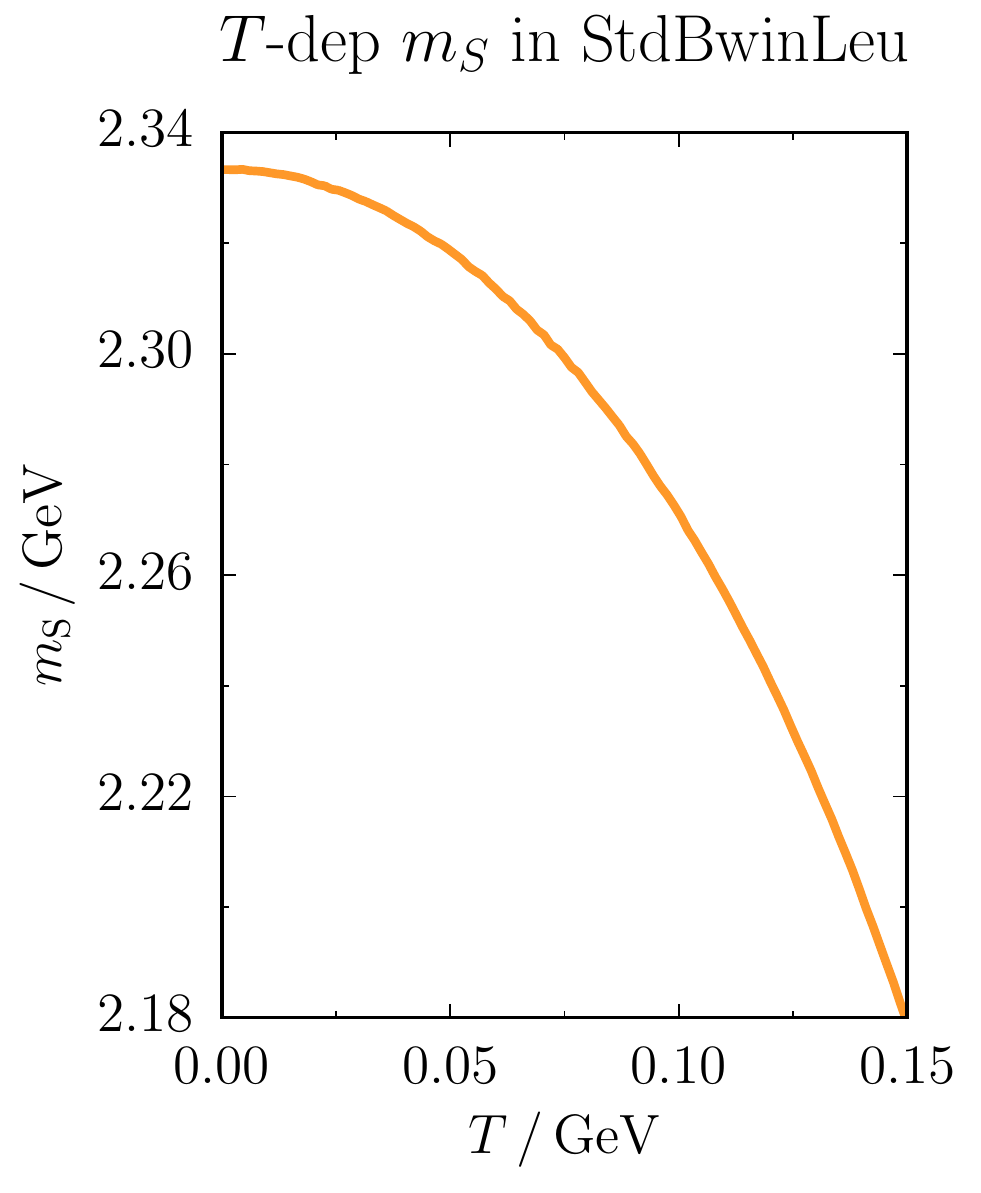}%
\includegraphics[trim=1.5mm 0mm 1.2mm 10mm,clip,height=0.98\imageheight]{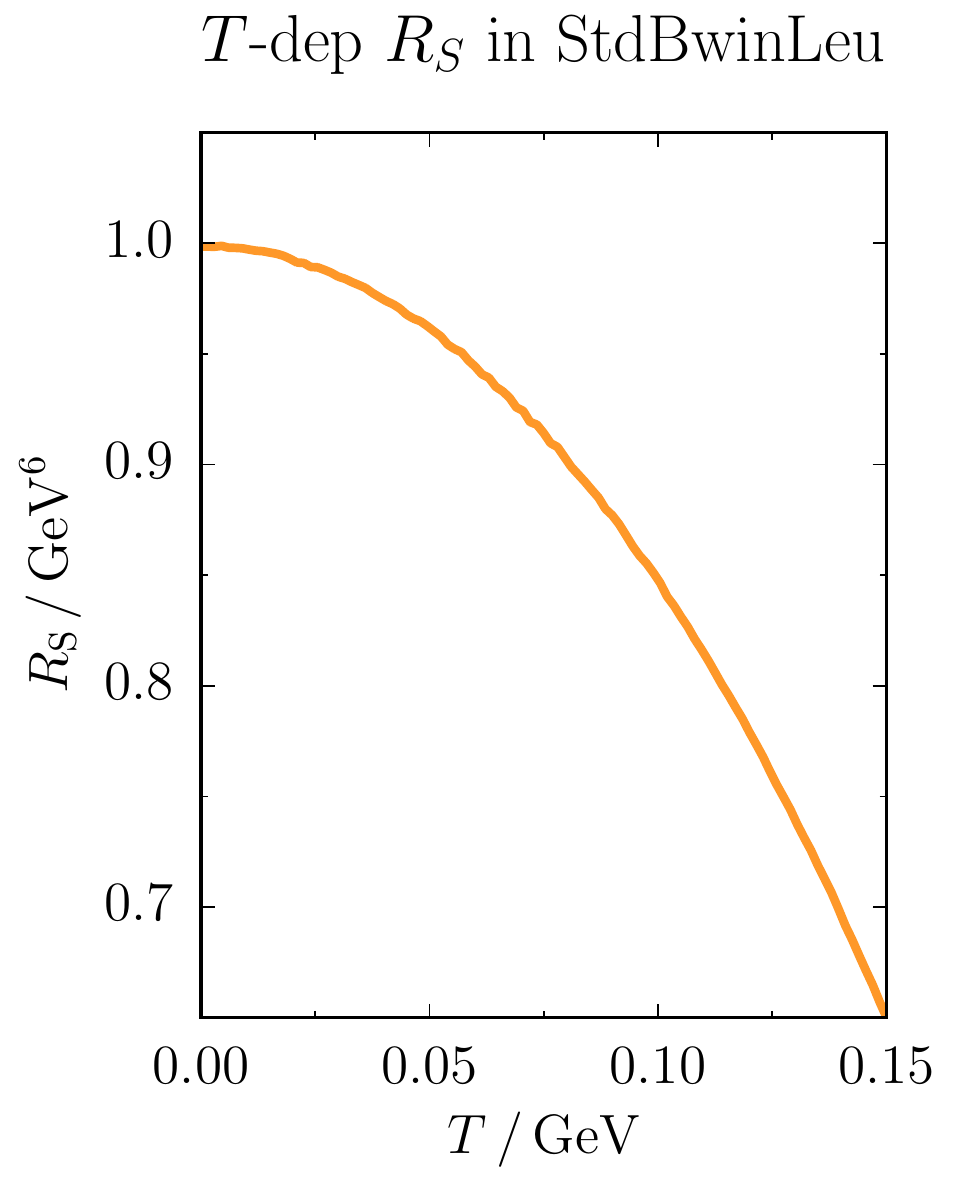}%
\includegraphics[trim=1.5mm 0mm 1.2mm 10mm,clip,height=0.98\imageheight]{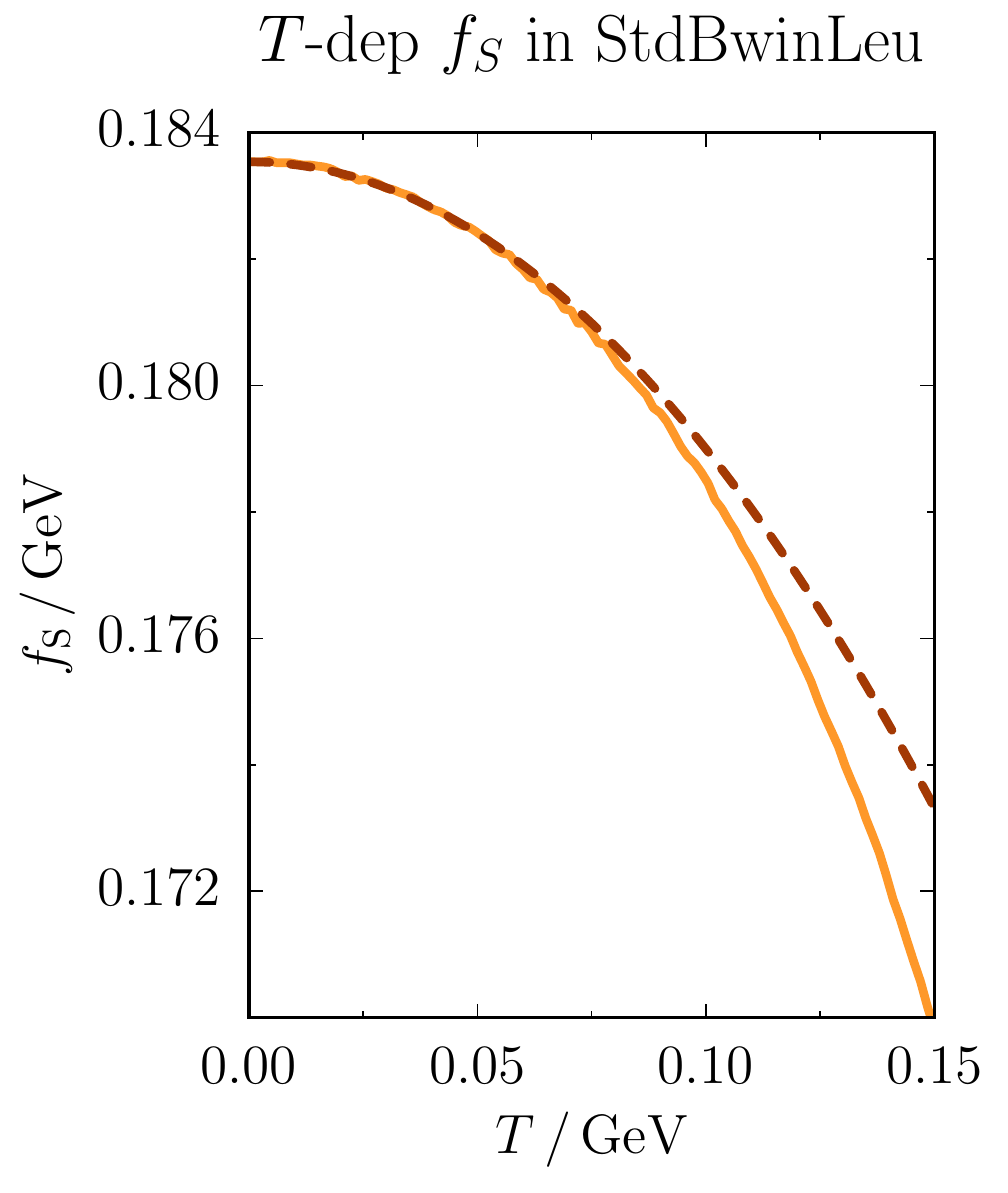}%
\caption[Temperature dependences of the spectral properties of the scalar D meson]{Temperature dependences of the spectral properties of the scalar D meson, where the left panel depicts the mass parameter $m_\mathrm{S}$, the center panel displays the residuum $R_\mathrm{S}$ and the right panel shows the decay constant $f_\mathrm{S}$ from our analysis (orange solid curve) as well as the temperature dependence of the decay constant deduced from Eq.~\eqref{eq:fTdepGassLeut} (brown dashed curve).}%
\label{fig:STdep}%
\end{figure}
A numerical Borel analysis for continuous temperatures up to $150\,\mathrm{MeV}$, where the low-temperature expansion of the condensates is assumed to hold \cite{Hatsuda:1992bv}, yields temperature dependent scalar D meson masses $m_\mathrm{S}$ and residua $R_\mathrm{S}$ averaged within the Borel window as depicted in Fig.~\ref{fig:STdep}.
These results exhibit a clear scalar D meson mass drop.
Provided the pseudo-scalar D meson mass is non-decreasing w.\,r.\,t.\ $T$, this points to approaching chiral partner meson masses for increasing temperatures, and thus, to chiral restoration.

The drop of $R_\mathrm{S}$ ($f_\mathrm{S}$) with increasing temperature indicates that the scalar D meson decouples from the scalar (vector) quark current.
That is a sign of deconfinement.
This interpretation is supported by the fact that the continuum threshold also drops with increasing temperature.
In other words, one sees the onset of the disappearance of the single-particle peak and its replacement by the quark--anti-quark continuum.
Thus, one sees precursors for both effects that are expected to happen in a hot and dense medium: chiral restoration and deconfinement.
What remains to be checked is whether {\em all} the properties of the scalar D meson move towards the properties of the pseudo-scalar partner.
We have seen the tendency for the masses.
How about the overlap with the quark currents?
Anticipating the results from Sec.~\ref{subsec:LuchaVacPS} we find $R_\mathrm{P} \approx 0.54\,\mathrm{GeV^6}$ and $f_\mathrm{P} \approx 0.21\,\mathrm{GeV}$ for the vacuum case.
We assume that this does not change much at low temperatures.
Comparison with the results of Fig.~\ref{fig:STdep} shows that $R_\mathrm{S}(T)$ drops very fast towards $R_\mathrm{P}(T=0)$, in qualitative agreement with expectations.
For $f_\mathrm{S}$, however, the situation is different.
Already its vacuum value is smaller than the pseudo-scalar decay constant and it drops further with temperature --- in agreement with the deconfinement picture.
Whether this apparent tension between chiral restoration and deconfinement is lifted by an in-medium drop of $f_\mathrm{P}$ remains to be seen, cf.\ Subsec.~\ref{subsec:LuchaMedS}.
We recall that the extraction of the in-medium (pseudo-)scalar decay constant is accompanied by some subtleties as already discussed below Eq.~\eqref{eq:fRrel}.
In any case, the more directly deduced overlap of the scalar meson with the scalar quark current shows the expected behavior of $R_\mathrm{S} \to R_\mathrm{P}$.

The temperature behavior of the scalar D meson mass $m_\mathrm{D}$ depicted in Fig.~\ref{fig:STdep} deviates from the result calculated from a hadronic approach incorporating heavy-quark symmetry as well as terms which explicitly break chiral symmetry \cite{Sasaki:2014asa}, cf.\  Fig.~\ref{fig:PSLuchasMed} below.
Although, both approaches predict a significant mass drop at high temperatures, the D meson mass deduced from the hadronic approach remains almost constant before dropping rapidly at $T\sim120\,\mathrm{MeV}$, while our \gls{QSR} evaluation points to a smooth (parabolic) temperature dependence.
Such a behavior is transferred to the respective residuum $R_\mathrm{S}$ and decay constant $f_\mathrm{S}$ which approximately obeys
\begin{align}\label{eq:fTdepGassLeut}
	f_\mathrm{S}(T) = f_\mathrm{S}(0) \left( 1 - \frac{T^2}{12 [f_\mathrm{S}(0)]^2} \right) \, ,
\end{align}
cf.\ Fig.~\ref{fig:STdep}.
In the framework of chiral perturbation theory, this functional form has been distilled from the leading order temperature dependence of the axial-vector correlator at large space-like momenta determining the pion decay constant $f_\pi$, where massless pions have been assumed, i.\,e.\ a two-flavor system which imposes the factor $1/12$ \cite{Gasser:1986vb}.
It is not clear whether this is just a coincidence or whether this points to a deeper relation between the in-medium behavior of the decay constants of different spin-0 mesons.
It would be also interesting to contrast these findings with results for $f_\mathrm{S}(T)$ in the hadronic approach of \cite{Sasaki:2014asa}.
This has not been calculated yet.

\subsection{Origin of temperature effects in both channels}
\label{subsubsec:originTeff}

Naively, the different temperature behaviors of the chiral partners may be attributed to canceling temperature dependences of the condensates in the pseudo-scalar \gls{OPE} contrasted by accumulating temperature dependent contributions in the scalar \gls{OPE}, i.\,e.\
\begin{align}
	\widetilde\Pi_\mathrm{P,S} (M^2;T) = \Pi_0^\mathrm{P,S} (M^2) + \Pi_T^\mathrm{P,S} (M^2)
\end{align}
with
\begin{subequations}
\begin{align}
	\Pi_0^\mathrm{P,S} (M^2) & = \frac{1}{\pi} \int\limits_{m_Q^2}^{s_0^\mathrm{P,S}} \rmd\omega \; e^{-\omega/M^2} \mathrm{Im} \Pi^\mathrm{pert}(\omega) + e^{-m_Q^2/M^2} m_Q^2 \Bigg[ \mp m_Q \langle \bar q q \rangle_0 \nonumber\allowdisplaybreaks\\
	& \phantom{=}  + \frac{1}{12}\langle \frac{\alpha_\mathrm{s}}{\pi} G^2 \rangle_0 \pm \frac{1}{2} \bigg( \frac{m_Q^3}{2M^4} - \frac{m_Q}{M^2} \bigg)\! \langle \bar q g \sigma G q \rangle_0 \Bigg] \: , \allowdisplaybreaks\\[1mm]
	\Pi_T^\mathrm{P,S} (M^2) & = T^2 B_1\!\!\left( \frac{m_\pi}{T} \right) e^{-m_Q^2/M^2} m_Q^2 \Bigg[ \pm \frac{m_Q}{8f_\pi^2} \langle \bar q q \rangle_0  + \frac{m_\pi^2}{18} + \frac{1}{4}\! \left( \frac{m_Q^2}{M^2} - 1 \right) \!\!\cdot 0.916 \nonumber\allowdisplaybreaks\\
	& \phantom{=} \times \left( \frac{\pi^2}{5}T^2 \frac{B_2\!\left(\frac{m_\pi}{T}\right)}{B_1\!\left(\frac{m_\pi}{T}\right)} - \frac{m_\pi^2}{8} \right) \mp \frac{1}{2} \bigg( \frac{m_Q^3}{2M^4} - \frac{m_Q}{M^2} \bigg) \frac{1}{8f_\pi^2} \langle \bar q g \sigma G q \rangle_0 \Bigg] \: ,
	\label{eq:OPE_Tdep}
\end{align}
\end{subequations}
cf.\ Tab.~\ref{tab:condTdep}.
In the light chiral limit, $m_\pi \rightarrow 0$, the only relevant terms at low temperatures in Eq.~\eqref{eq:OPE_Tdep} arise from the chirally odd contributions, i.\,e.\ the chiral condensate and the mixed quark-gluon condensate.
Hence, no significant cancelation of channel-specific  and channel-independent terms occurs, but the relevant temperature dependent terms enter the pseudo-scalar and scalar \glspl{OPE} $\widetilde\Pi_\mathrm{P,S}$ with opposite sign.
Since the different temperature behaviors of pseudo-scalar and scalar mesons cannot be understood by comparing the temperature dependent \gls{OPE} contributions $\Pi_T^\mathrm{P,S}$, one needs to disclose intermediate steps of the Borel analysis to explain the phenomenon in the scope of \glspl{QSR}.

We recall that, in this framework, meson masses are evaluated as the average of the particular meson mass Borel curve, which shows maximum flatness in the corresponding Borel window.
The meson mass Borel curves of pseudo-scalar and scalar D mesons feature a pole below the Borel window, which lifts the mass Borel curve in the Borel window and subsequently increases the meson mass average.
These poles originate from zeros of the \glspl{OPE} $\widetilde\Pi_\mathrm{P,S} (M)$, entering the denominator of the mass Borel curve formula~\eqref{eq:ratioQSR}, and are subject to changes at higher temperatures which turn out to be very different for pseudo-scalar and scalar mesons, cf. Appendix~\ref{app:TdepOPE} for further details.

While the pole in the pseudo-scalar curve is hardly shifted, the relevant pole of the scalar meson mass Borel curve $m_\mathrm{S}(M)$ experiences sizable shifts to lower Borel masses for growing temperatures and vanishes at high $T$.
Due to the vicinity of the Borel window the scalar D meson mass is effected by such a drastic temperature behavior of the pole structure of the scalar meson mass Borel curve.
Although the Borel window boundary $M_\mathrm{min}^\mathrm{S}$ also moves to lower Borel masses for rising temperatures, the minimum of the mass Borel curve at the vacuum value of the threshold $s_0^\mathrm{S} \sim 8\,\mathrm{GeV}^2$ drifts to the left boundary of the Borel window due to the ($s_0^\mathrm{S}$-independent) pole shift.
Thus, the flatness requirement of the mass Borel curve, where the minimum of this curve is approximately centered within the Borel window, is met at lower values of the continuum threshold parameter $s_0^\mathrm{S}$, cf.\ Fig.~\ref{fig:Sm(M,s)cuts}.
This mechanism, inherent to the scalar D meson sum rule, causes the enhanced temperature effects which superimpose the modest temperature modifications of the masses and residua in wide sections within the relevant Borel mass range as depicted in Fig.~\ref{fig:PSmR(M,s)dT}.

\section{Borel analysis with given meson mass input}
\label{sec:BAgivInput}

The conventional Borel analysis as performed in the previous section successfully provides temperature dependent spectral information for the scalar D meson, whereas the equivalent data can not be reliably deduced for its pseudo-scalar partner.
In order to obtain these results, the conventional analysis is contrasted by an analysis which utilizes the given meson mass as input and aims for deducing the hadron's residuum and decay constant, already successfully applied to pseudo-scalar D mesons in vacuum \cite{Lucha:2010ea,Lucha:2011zp}.
This optimized approach utilizes a Borel mass dependent continuum threshold ansatz $s_0(M) = \sum_{n=0}^{n_\mathrm{max}} \frac{s_{(n)}}{M^{2n}}$ to obtain unbiased mesonic decay constants, cf.\ Appendix~\ref{app:luchaQSR} for details.

\subsection[Vacuum Borel curves and extraction of |V\_cd|]{Vacuum Borel curves and extraction of $\boldsymbol{|V_{cd}|}$}
\label{subsec:LuchaVacPS}

\begin{figure}[t]
\settoheight{\imageheight}{\includegraphics[trim=0mm 0mm 0mm 15mm,clip,width=0.49\columnwidth]{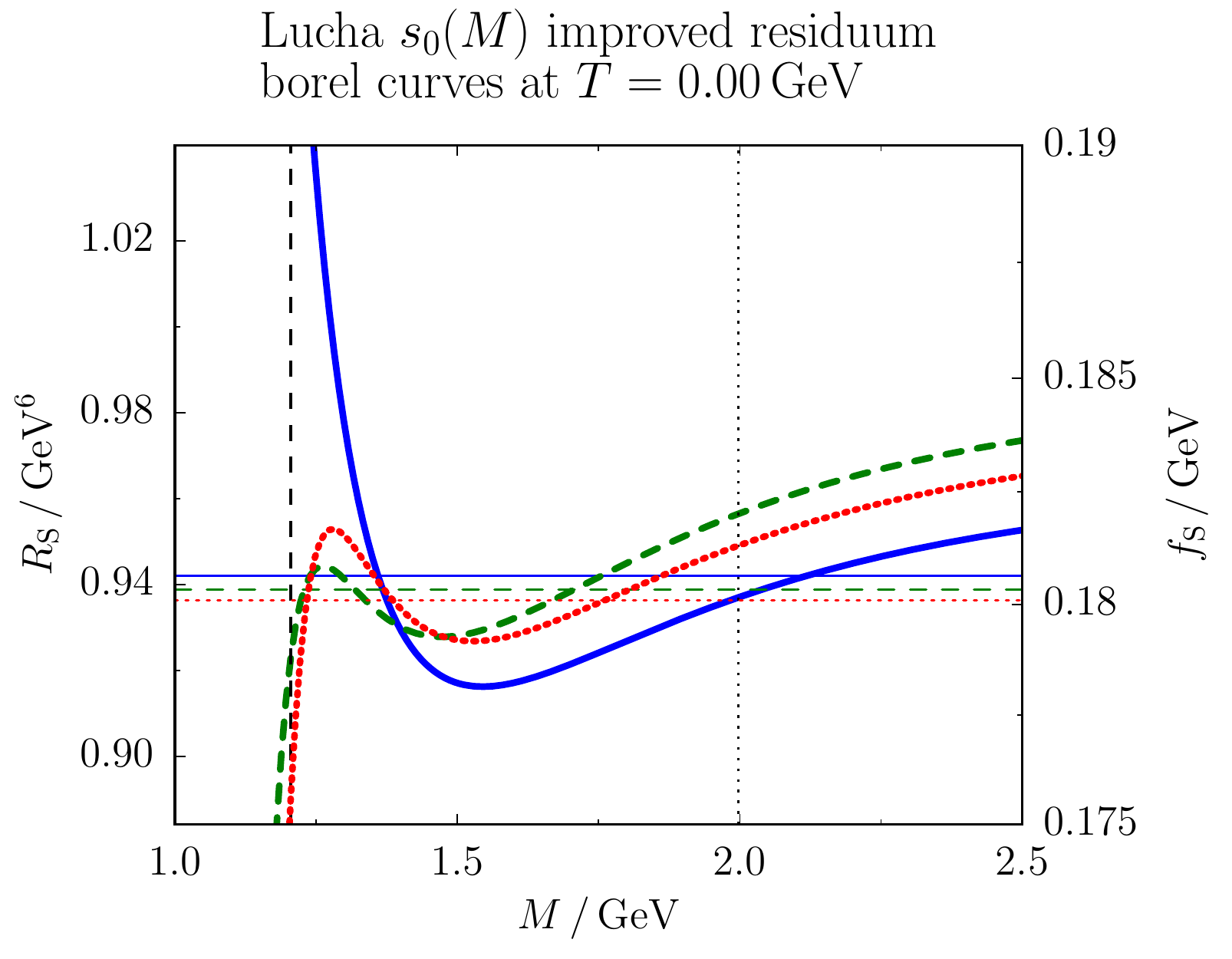}}
\begin{tabular}{@{}lp{0.5mm}l@{}}
\includegraphics[trim=2.0mm 0mm 0mm 15mm,clip,height=\imageheight]{fig7a.pdf}%
&&
\includegraphics[trim=0mm 0mm 2.5mm 15mm,clip,height=\imageheight]{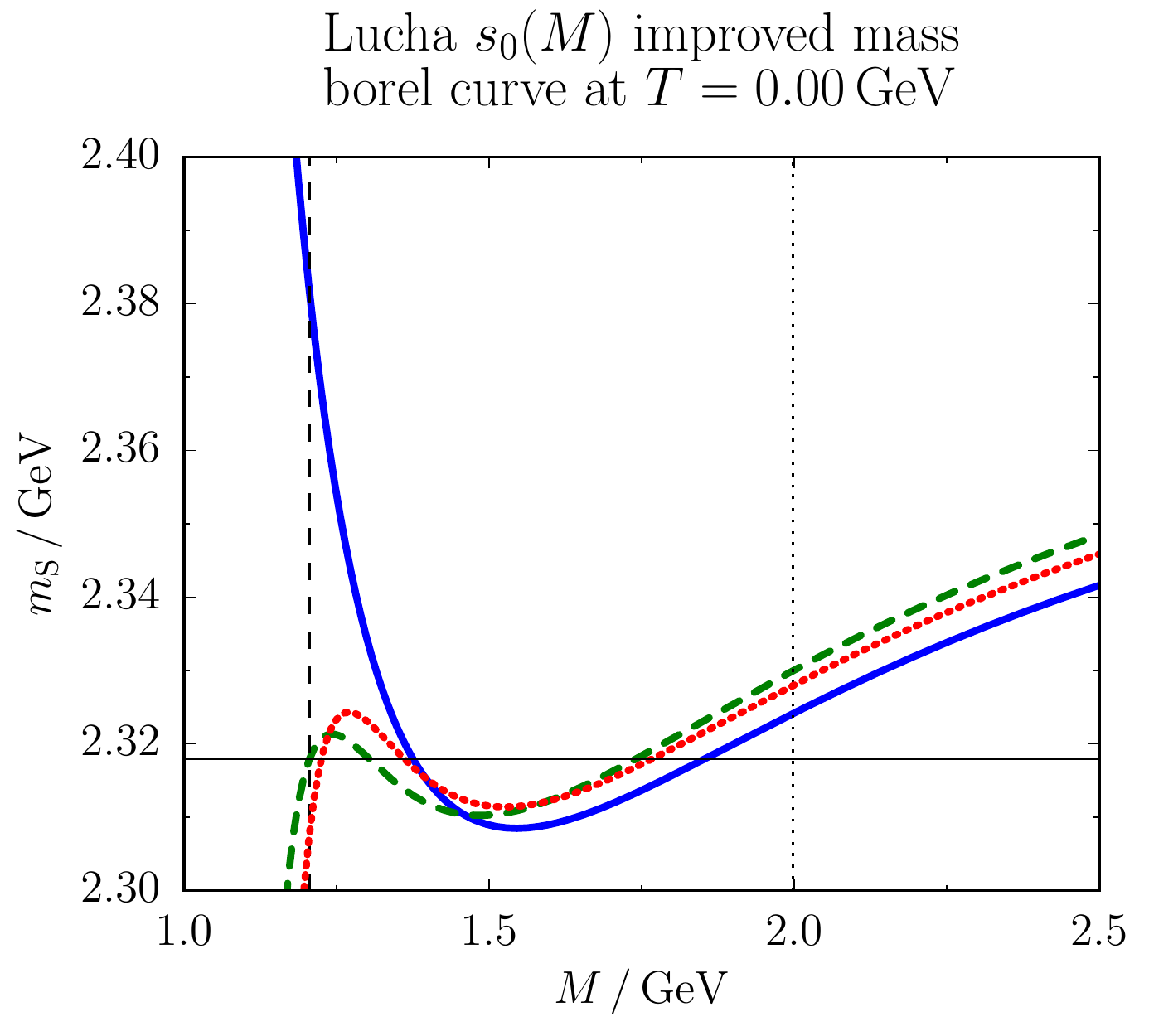}%
\\
\includegraphics[trim=2.8mm 0mm 0mm 15mm,clip,height=\imageheight]{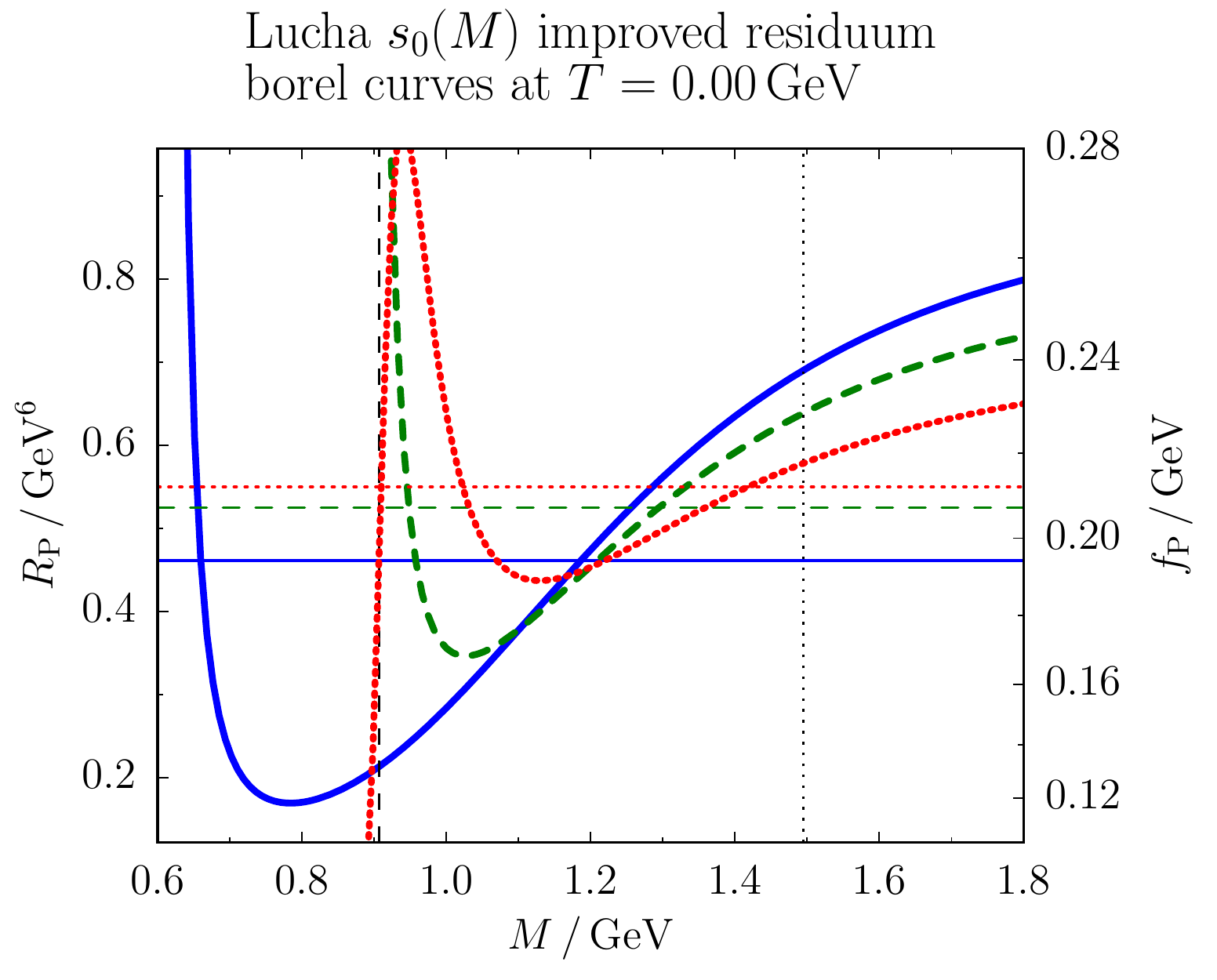}%
&&
\includegraphics[trim=0mm 0mm 2.5mm 15mm,clip,height=\imageheight]{fig7d.pdf}%
\end{tabular}
\caption[Pseudo-scalar and scalar D meson Borel curves optimized with $M$-dependent continuum threshold]{%
Pseudo-scalar (left panels) and scalar (right panels) D meson Borel curves in vacuum.
The upper panels display the mass Borel curves for $M$-dependent continuum threshold parameters~\eqref{eq:defLuchas} with minimized deviations from the actual meson mass depicted by the solid horizontal line.
The blue solid, green dashed and red dotted curves correspond to continuum thresholds with the degree of the polynomial ansatz $n_\mathrm{max}=0$, 1 and 2, respectively. The residuum and decay constant Borel curves depicted in the lower panels are associated with the mass Borel curves in the upper panels, i.\,e.\ the same color code applies, where the corresponding horizontal lines depict the Borel window average.
The vertical dashed and dotted lines denote the respective lower and upper Borel window boundaries.}%
\label{fig:PSLuchasVac}%
\end{figure}

Before deducing temperature dependent decay properties, we study the residua and decay constants of the D mesons in vacuum at $T=0$ to test this approach and to extract the off-diagonal CKM matrix element $|V_{cd}|$.
Our evaluations show that the minimization procedure incorporated into the approach yields acceptable results only for a fixed Borel window and fails for continuum threshold parameter dependent upper Borel window boundaries $M_\mathrm{max}(s_0)$.
To reduce the impact of the continuum on the spectral properties of the pseudo-scalar and scalar D mesons the Borel windows from the conventional analysis, $[0.9\,\mathrm{GeV},1.5\,\mathrm{GeV}]$ and $[1.2\,\mathrm{GeV},2\,\mathrm{GeV}]$, respectively, cf.\ Fig.~\ref{fig:PSm(M,s)StdBwin} upper panels, have been utilized to obtain the results in Fig.~\ref{fig:PSLuchasVac}.

The mass Borel curves (upper panels) for an increasing number of polynomials contributing to the $M$-dependent continuum threshold parameter adapt closer to the actual D meson mass in both channels.
We recover previous findings for the pseudo-scalar decay constant in the range of $f_\mathrm{P}=(201-211)\,\mathrm{MeV}$ in Ref.~\cite{Gelhausen:2013wia,Narison:2012xy,Lucha:2011zp,Wang:2013ff} as well as the tendency of rising residuum and decay constant values for increasing $n_\mathrm{max}$ reported in \cite{Lucha:2010ea} for pseudo-scalar D mesons, cf.\ lower left panel of Fig.~\ref{fig:PSLuchasVac}.%
\footnote{%
However, mind that we produce Borel curves up to $n_\mathrm{max}=2$ in contrast to Ref.~\cite{Lucha:2010ea} containing also results with $n_\mathrm{max}=3$.
Due to optimization parameters $s_{(n)}$ changing their order of magnitude for $n_\mathrm{max}=3$ compared to $n_\mathrm{max}=1$ and $2$ we disregard these results.
This phenomenon may be attributed to our Borel window which is rigorously determined from standard requirements \cite{Leinweber:1995fn} but differs from the one in Ref.~\cite{Lucha:2010ea}, because this numerical discrepancy is gradually lifted when shifting the Borel window to higher Borel masses $M$.%
}
While vacuum scalar D meson decay constants deduced from a \gls{QSR} with perturbative term in order $\alpha_\mathrm{s}^2$ are reported to reside in the range of $f_\mathrm{S}=(217-221)\,\mathrm{MeV}$ in Refs.~\cite{Narison:2003td,Narison:2015nxh}, the determination \cite{Colangelo:1991ug} building on the perturbative term in order $\alpha_\mathrm{s}$, used throughout this work, yields $f_\mathrm{S} = 170\,\mathrm{MeV}$ which is relatively close to our findings depicted in the right panels of Figs.~\ref{fig:PSLuchasVac} and \ref{fig:PSLuchasMed}, cf.\ the consistent (vacuum) result in the right panel of Fig.~\ref{fig:STdep}.
As our pseudo-scalar and scalar D meson results resemble previous vacuum results we apply this approach to pseudo-scalar and \mbox{scalar D mesons at finite temperatures in the next subsection.}

In order to determine the off-diagonal CKM matrix element, $|V_{cd}|$, by virtue of Eq.~\eqref{eq:width}, the pseudo-scalar D meson has to be used, because the branching fraction for the needed decay $\mathrm{S^+}\rightarrow\ell^+\nu_\ell$ is not available.
Apart from the decay constant $f_\mathrm{P}$ all necessary numerical values can be found in Ref.~\cite{Olive:2016xmw}.
The Borel curves for extracting $f_\mathrm{P}$ are shown in Fig.~\ref{fig:PSLuchasVac}, where the averaged decay constants (depicted by horizontal lines in the lower left panel) are $f_\mathrm{P}=194$, 207 and $212\,\mathrm{MeV}$ for $n_\mathrm{max}=0$, 1 and 2, respectively.
Depending on the degree of the polynomial continuum threshold ansatz the resulting CKM matrix element varies between $|V_{cd}|=0.219$ and $0.239$ which is in agreement with $|V_{cd}|^\text{PDG}=0.230 \pm 0.011$ from Ref.~\cite{Olive:2016xmw}.
Employing the decay constant with $n_\mathrm{max}=2$, providing presumably the most reliable numerical value, one obtains $|V_{cd}|=0.219$ for the CKM matrix element with $\sim10\,\mathrm{\%}$ uncertainty, where aside from the branching fraction measurement, the extraction of the decay constant gives the largest relative uncertainty if one takes the decay constant range from $n_\mathrm{max}=0$ to 2 as a rough estimate.

\subsection{Temperature effects}
\label{subsec:LuchaMedS}

Extracting the decay constant from a fixed meson mass by adjusting multiple coefficients of a Borel mass dependent continuum threshold is also viable in a strongly interacting medium if the temperature and/or net-baryon density dependence of the respective meson mass is at our disposal.
The resulting decay properties of chiral partner mesons may contain signals of (partial) chiral restoration in the medium.

\begin{figure}[t]
\centering
\settoheight{\imageheight}{\includegraphics[trim=0mm 0mm 0mm 18mm,clip,width=0.49\columnwidth]{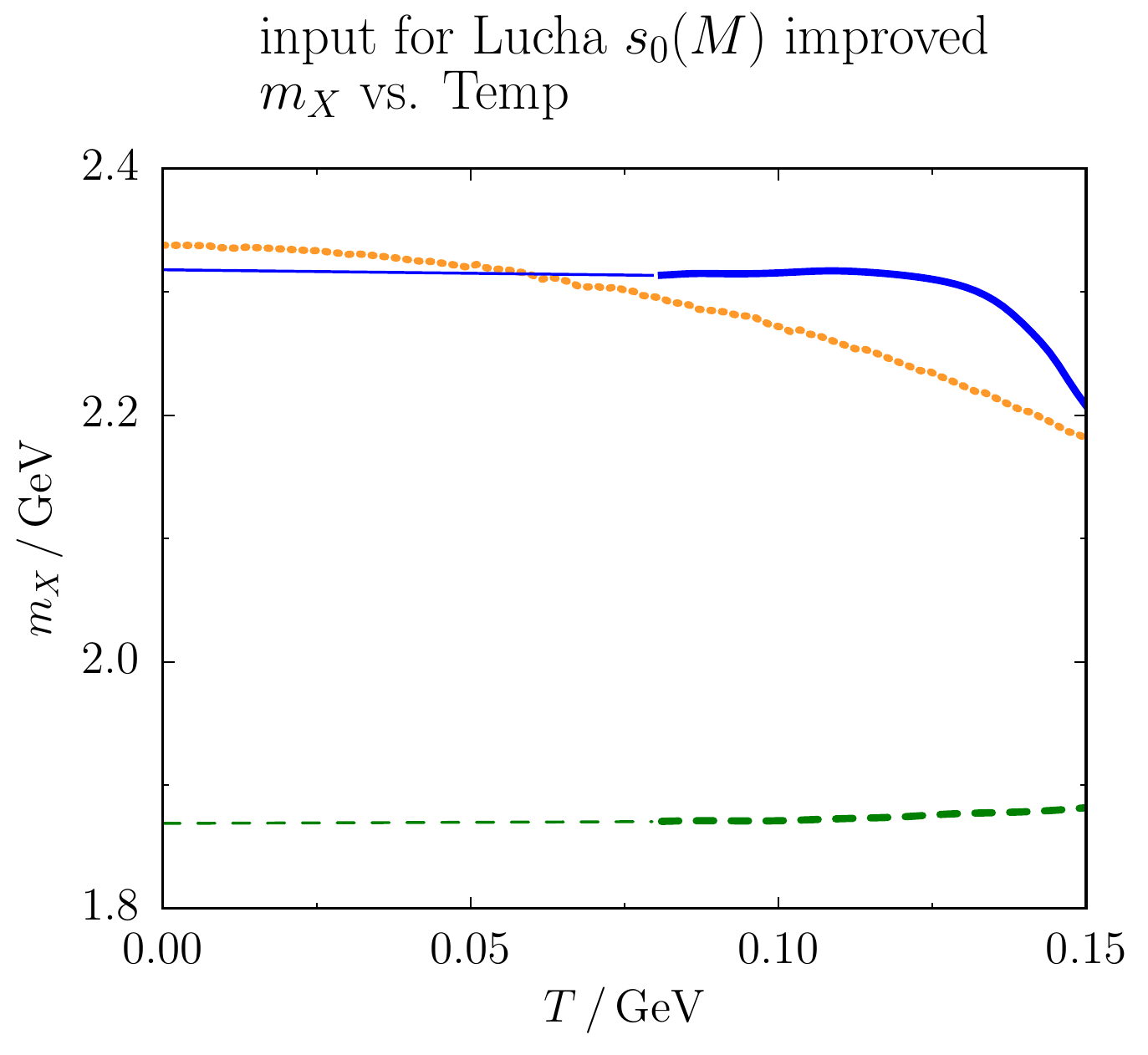}}
\includegraphics[trim=1.5mm 0mm 1.2mm 17mm,clip,height=\imageheight]{fig8.pdf}%
\caption[Temperature dependent input]{Temperature dependences of the input mass parameters for pseudo-scalar and scalar D mesons used for the optimized sum rule evaluation in this subsection.
The pseudo-scalar and scalar D meson mass parameters calculated in Ref.~\cite{Sasaki:2014asa} are depicted by the thick green dashed and thick blue solid curves, respectively.
The thin curves of the same color are their linear extensions to $T=0$.
The orange dotted curve shows the temperature dependence of the scalar D meson mass extracted with the conventional approach, cf.\ left panel of Fig.~\ref{fig:STdep}.}%
\label{fig:PSLuchasMedInput}%
\end{figure}

Due to lacking experimental information we use the temperature dependent D meson masses that are calculated within a hadronic approach comprising chiral symmetry breaking terms \cite{Sasaki:2014asa} as well as the scalar D meson mass results from Sec.~\ref{subsubsec:numS} for a comparison, cf.\ Fig.~\ref{fig:PSLuchasMedInput}.
As the temperature curves in Ref.~\cite{Sasaki:2014asa} are given in the temperature range of $T=(80-230)\,\mathrm{MeV}$ we extrapolate these mass curves to facilitate our low-temperature evaluation covering $T=(0-150)\,\mathrm{MeV}$.
Starting from the pseudo-scalar and scalar D meson vacuum values \cite{Olive:2016xmw} the mass curves are connected to the associated curves from \cite{Sasaki:2014asa} by a straight line. 

\begin{figure}[t]
\centering
\settoheight{\imageheight}{\includegraphics[trim=0mm 0mm 0mm 18mm,clip,width=0.49\columnwidth]{fig8.pdf}}
\includegraphics[trim=1.5mm 0mm 1.2mm 17mm,clip,height=\imageheight]{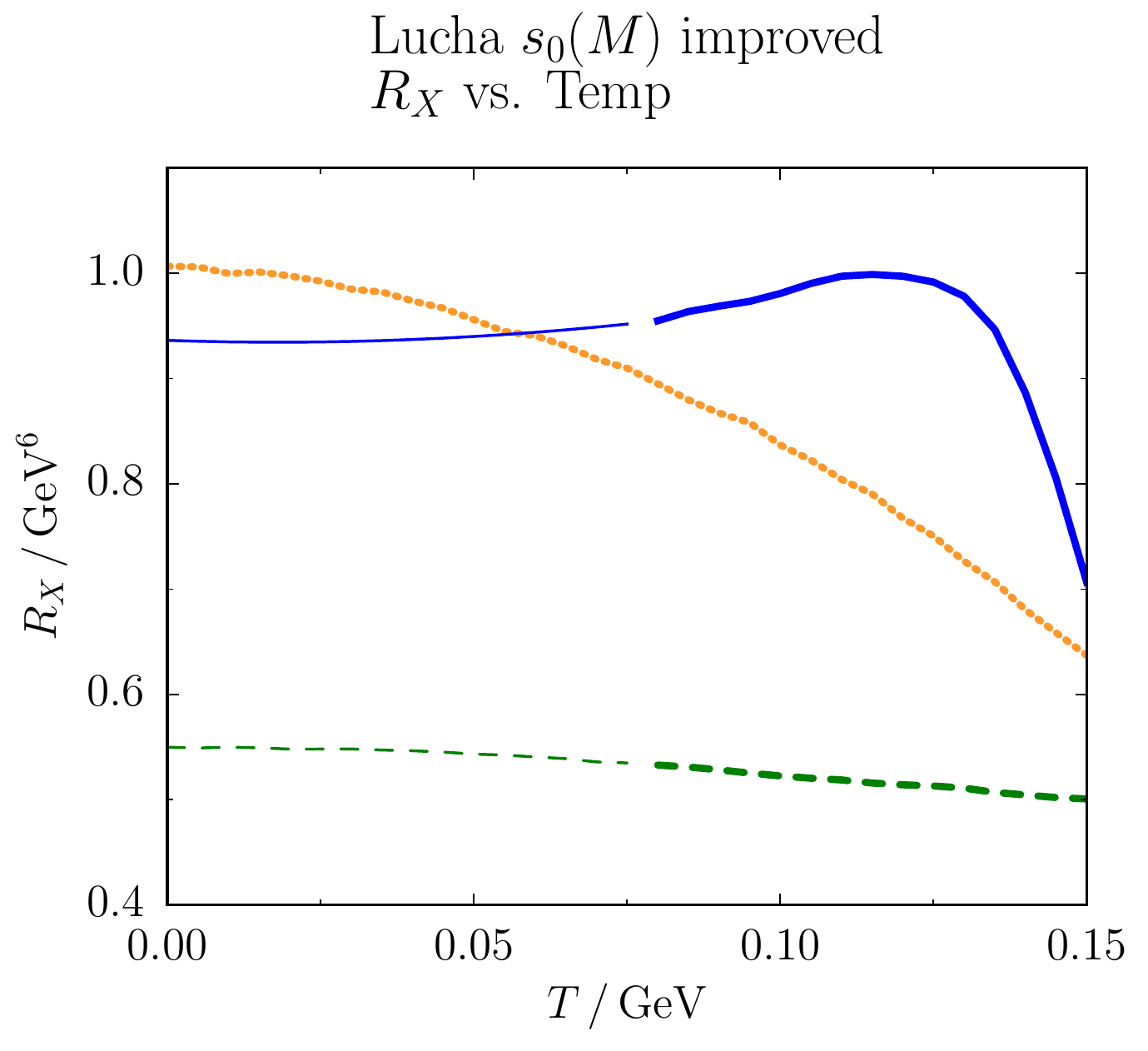}%
\hfill
\includegraphics[trim=1.5mm 0mm 1.2mm 17mm,clip,height=\imageheight]{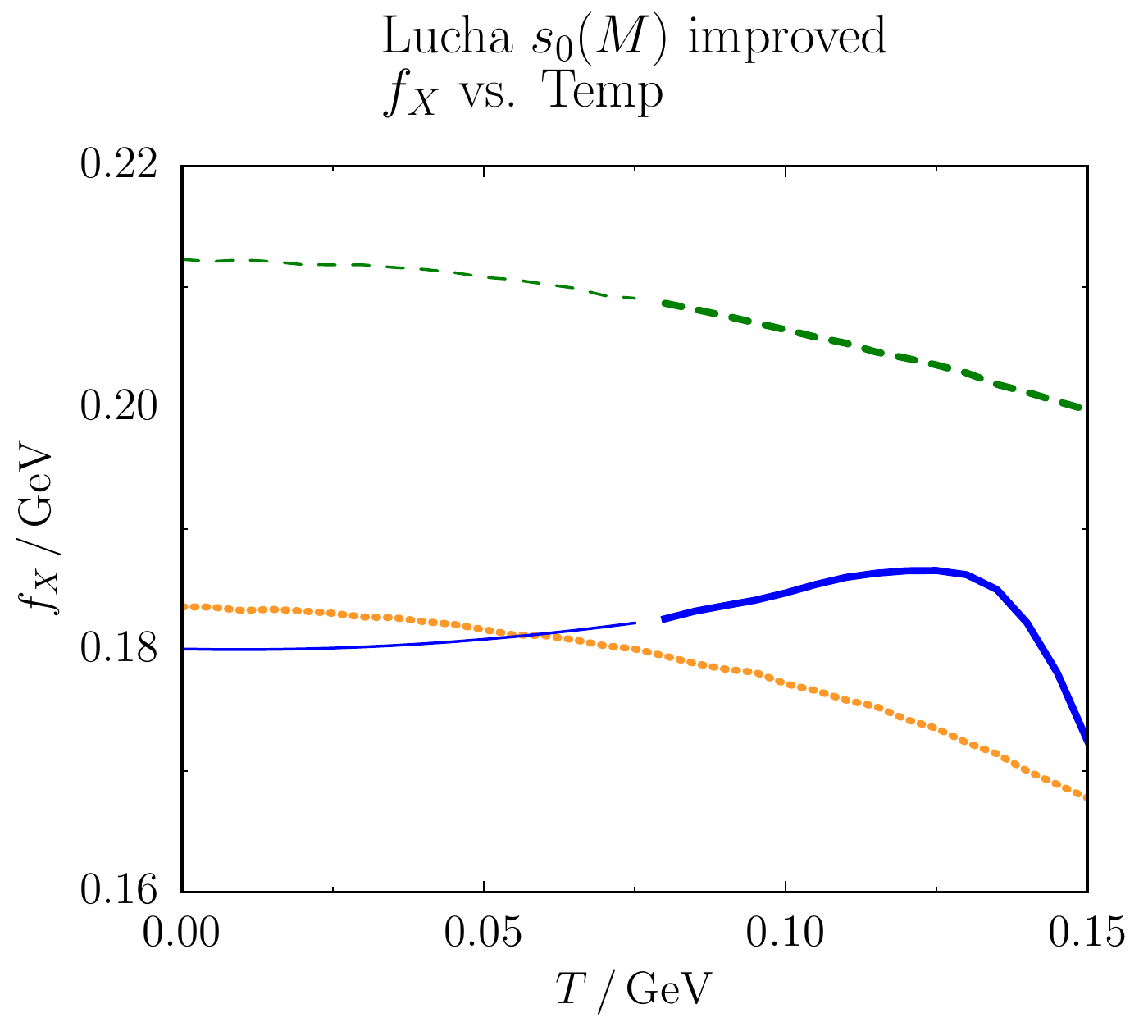}%
\caption[Temperature dependences resulting decay properties]{Temperature dependences of the decay properties encoded in $R_X$ (left) and $f_X$ which are based on the input in Fig.~\ref{fig:PSLuchasMedInput}, where the same color code applies.
The depicted results are computed from a continuum threshold with second degree of the polynomial ansatz, i.\,e.\ $n_\mathrm{max}=2$.}%
\label{fig:PSLuchasMed}%
\end{figure}

We studied various Borel window configurations building on the vacuum and $T=150\,\mathrm{MeV}$ Borel windows from the 'conventional analysis',%
\footnote{In accordance with the vacuum evaluation, $s_0$-dependent upper Borel boundaries $M_\mathrm{max}(s_0)$ obstruct the optimization procedure also at finite temperatures.
Hence, a fixed Borel window has to be deployed for each temperature.
The Borel windows from the conventional analysis can provide a rough estimate only, because they refer to an optimized continuum threshold parameter at a given temperature.%
} 
cf.\ Fig.~\ref{fig:Sm(M,s)cuts}, e.\,g.\ temperature independent intersection and union of the respective Borel windows, or a simple construction where the lower boundary changes linearly from its vacuum value $M_\mathrm{min}(T=0)$ to $M_\mathrm{min}(T=150\,\mathrm{MeV})$ and $M_\mathrm{max}$ alike.
The extracted numerical values are sen\-si\-tive to the definition of the temperature dependent Borel window or even produce implausible optimization results.
Reliable numerical results, displayed in Fig.~\ref{fig:PSLuchasMed}, are obtained with the temperature independent Borel window adopted from the evaluation in vacuum.

While we inferred a negligible temperature dependence of the pseudo-scalar D meson mass from Fig.~\ref{fig:PSmR(M,s)dT} that could not be quantified in the conventional way, we here use a slightly growing D meson mass calculated in Ref.~\cite{Sasaki:2014asa} to obtain the pseudo-scalar decay properties.
As anticipated in the previous section the residuum $R_\mathrm{P}(T)$ undergoes only subtle changes depicted by the green dashed curve in the left panel of Fig.~\ref{fig:PSLuchasMed}.
It mildly drops in agreement with the expectation for the pseudo-scalar meson quark current overlap $R_\mathrm{P} = |\langle 0 | j_\mathrm{P}(0) | \mathrm{P} \rangle|^2$ as a precursor of deconfinement in a strongly interacting medium.
The corresponding decay constant $f_\mathrm{P}(T)$ also decreases, cf.\ Fig.~\ref{fig:PSLuchasMed} right panel.

For growing temperatures the residuum $R_\mathrm{S}(T)$, obtained with input from Ref.~\cite{Sasaki:2014asa}, i.\,e.\ the blue solid curve, increases by $~5\,\mathrm{\%}$ at $T=120\,\mathrm{MeV}$ before dropping rapidly.
The associated temperature curve of the decay constant $f_\mathrm{S}(T)$ exhibits a similar behavior.
The rising $R_\mathrm{S}(T)$ and $f_\mathrm{S}(T)$ at low temperatures are driven by the rapidly changing chiral condensate while the scalar meson mass remains almost constant.
When $m_\mathrm{S}$ decreases at higher values of $T$ it dominates the temperature behavior of the scalar residuum and decay constant, yielding rapidly dropping temperature curves.
The final decrease fits to the chiral restoration scenario.
Whether the small intermediate rise is of physical significance or an artifact of the hadronic-model input or lies numerically inside of the intrinsic uncertainty of the QSR method remains to be seen.

The resulting temperature dependences of $R_\mathrm{S}(T)$ and $f_\mathrm{S}(T)$ obtained with the scalar D meson mass input from the conventional analysis, depicted by the orange dotted curves, exhibit the expected qualitative behaviors right away.
As the input mass drops fast enough for rising temperatures, the residuum and decay constant decrease monotonically.
Signals of chiral symmetry restoration can be seen in the left panel of Fig.~\ref{fig:PSLuchasMed}, where the pseudo-scalar (green dashed) and scalar (orange dotted) residuum curves approach each other at high temperatures, whereas the decay constant curves in the right panel exhibit an approximately constant gap.
However, due to the relation \eqref{eq:fRrel} between residuum and decay constant it is a matter of the temperature dependence of the meson mass if precursors of chiral symmetry restoration also translate to $f_\mathrm{S}(T)$.
If the meson mass decreases slower than the orange dotted curve in Fig.~\ref{fig:PSLuchasMedInput}, but fast enough to ensure a decreasing residuum, $f_\mathrm{S}(T)$ may decrease slower than $f_\mathrm{P}(T)$ signaling partial chiral symmetry restoration, as well.

As anticipated in Subsec.~\ref{subsubsec:numS}, the tension between chiral symmetry restoration and deconfinement is relieved, indeed, because both types of spectral parameters, $R_X(T)$ and $f_X(T)$, can decrease while the individual curves of chiral partner D mesons approach each other, simultaneously.
This evaluation of (partial) chiral symmetry restoration patterns of parameters relevant to deconfinement effects exemplifies that \gls{DCSB} and confinement can not be studied separately but their relations \cite{Suganuma:2017syi} are to be taken into account.

\section{Summary}
\label{sec:sum}

Besides confinement, chiral symmetry breaking is the central phenomenon of {QCD} because it provides a mass generating mechanism giving essentially mass to the light hadrons.
As this mechanism is based on a spontaneous symmetry breaking principle the chiral symmetry breaking pattern as well as its restoration in a strongly interacting medium are subject to a large variety of investigations.
While previous studies often consider chiral effects on light mesons \cite{Dominguez:1989bz,Dominguez:2003dr,Dominguez:2014fua,Kapusta:1993hq,Kwon:2008vq,Kwon:2010fw,Hohler:2012xd,Hohler:2013eba,Ayala:2014rka}, we shift the focus to the heavy-light sector evaluating pseudo-scalar and scalar D meson \glspl{QSR}, because the notions of chiral symmetry 
can be translated into the heavy-light sector supposed the symmetry transformations are restricted to the light-quark content.
While pseudo-scalar D mesons have already been investigated in the framework of \glspl{QSR} in the vacuum \cite{Aliev:1983ra,Lucha:2011zp,Narison:2012xy} and in the medium \cite{Hayashigaki:2000es,Hilger:2008jg,Suzuki:2015est,Wang:2016rac}, the investigations in the present paper provide \gls{QSR} results in vacuum and at finite temperatures for pseudo-scalar as well as scalar D mesons, allowing for insights into the \gls{DCSB} phenomenology.

While the conventional \gls{QSR} analysis is inadequate to extract the pseudo-scalar {D} meson mass, its scalar counterpart can be treated successfully.
However, from intermediate steps of the analysis a particular insensitivity of the pseudo-scalar \gls{QSR} to temperature changes is evident, suggesting a negligible modification of the pseudo-scalar D meson spectral properties.
As the scalar \gls{QSR} evaluation yields a decreasing mass for growing temperatures, the channel-specific chiral partner sum rules signal the onset of partial chiral restoration.
We regard our low-temperature approximation valid up to about $150\,\mathrm{MeV}$.
This approximation is necessary for a reliable and model-independent evaluation of the condensates in a hadronic thermal medium.
The behavior of constant pseudo-scalar D mass and dropping scalar D mass is in qualitative agreement with the findings of Ref.~\cite{Sasaki:2014asa}, but the parabolic temperature curve deviates from the scalar D meson mass curve presented there.

Although, medium modifications of the {D} meson masses do not lead to measurable changes of the {D} meson production in a statistical hadronization model \cite{Andronic:2007zu}, the D meson yields in heavy-ion collisions may be sensitive to their altered decay properties in an ambient strongly interacting medium.
Accordingly, we have extracted the temperature dependences of the scalar {D} meson decay constants utilizing channel-specific \glspl{QSR}.
Due to their connection to particular leptonic branching fractions such decay constants are of large interest allowing for the determination of the off-diagonal {CKM} matrix element $|V_{cd}|=0.219$ at $T=0$ as a bonus.

The growing interest in decay constants of open charm mesons has lead to \glspl{QSR} for these quantities using the experimentally determined vacuum masses as phenomenological input \cite{Narison:2001pu,Lucha:2011zp}.
Hence, employing the estimated temperature behavior of these masses~\cite{Sasaki:2014asa} allows for the prediction of their in-medium decay constants.
An improvement of the \gls{QSR} analysis by introducing a Borel mass dependent continuum threshold parameter, which is supposed to suppress contaminations of the lowest resonance from continuum excitations of the spectral density, results in residuum and decay constant temperature curves deviating from the ones of the conventional analysis, i.\,e.\ albeit showing signals of chiral restoration at high temperatures the scalar residuum and decay constant do not decrease monotonically.
This tension between chiral symmetry restoration and deconfinement can be lifted, if scalar D meson mass curves are used which drop significantly already at low temperatures, e.\,g.\ the resulting mass temperature curve from the conventional analysis.

While the planned facilities at {NICA} \cite{Kekelidze:2016hhw}, {FAIR} \cite{Friman:2011zz} and {J-PARC} \cite{Sako:2014fha} will address charm degrees of freedom in a baryonic dense medium, the running collider experiments at {LHC} and {RHIC} are delivering at present a wealth of data on charm and bottom degrees of freedom in a high-temperature environment at very small net-baryon density.
The firm application of \glspl{QSR} on these quite different experimental conditions and the relation to observables, in particular those supporting the quest for chiral restoration signatures, deserve much more dedicated investigations on the theory side.

\section*{Acknowledgments}
The authors gratefully acknowledge enlightening discussions with S.\ J.\ Brodsky, S.\ H.\ Lee, K.\ Morita, U.\ Mosel, S.\ Narison, R.\ Rapp, R.\ Thomas, W.\ Weise and S.\ Zschocke.
We highly appreciate the conversation with W.\ Lucha on his approach in Ref.~\cite{Lucha:2011zp}.

\appendix

\section{Finite chiral transformations in the heavy-light sector}
\label{app:chitrafo}

General pseudo-scalar and scalar two-quark currents read
\begin{align}
	j_\mathrm{P}^\tau = i \bar \psi \gamma_5 \tau \psi \qquad \text{and} \qquad j_\mathrm{S}^\tau = \bar \psi \tau \psi
\end{align}
with $N_\mathrm{f}$-dimensional flavor vector $\psi$ and flavor matrix $\tau$.
They are decomposable into the (iso-vector) currents $j_\mathrm{P}^{(a)} = i\bar\psi\gamma_5(\tau^a)\psi$ and $j_\mathrm{S}^{(a)} = \bar\psi(\tau^a)\psi$, where the matrices $\tau^a$ acting on the flavor indices are the $N_\mathrm{f}^2-1$ traceless generators of $\mathrm{SU}(N_\mathrm{f})$.
One can rewrite these general currents using the decompostion of the flavor vector in left and right-handed parts: $\psi=\psi_\mathrm{L}+\psi_\mathrm{R}$ with $\psi_\mathrm{L,R}=P_\mathrm{L,R}\psi$ and the projectors $P_\mathrm{L,R}=(1\mp\gamma_5)/2$:
\begin{align}
	j_\mathrm{P}^\tau	& = i \bar \psi_\mathrm{L} \gamma_5 \tau \psi_\mathrm{R} + i \bar \psi_\mathrm{R} \gamma_5 \tau \psi_\mathrm{L}
	= \frac{i}{2}\left( \frac{1}{i} j_\mathrm{P}^\tau + \bar\psi_\mathrm{L}\tau\psi_\mathrm{R} - \bar\psi_\mathrm{R}\tau\psi_\mathrm{L} \right)
	= i\left( \bar\psi_\mathrm{L}\tau\psi_\mathrm{R} - \bar\psi_\mathrm{R}\tau\psi_\mathrm{L} \right) \nonumber\\
	& = i\left( j^\tau_\mathrm{LR} - j^\tau_\mathrm{RL} \right)
\end{align}
and
\begin{align}
	j_\mathrm{S}^\tau & = \bar \psi_\mathrm{L}\tau \psi_\mathrm{R} + \bar \psi_\mathrm{R} \tau \psi_\mathrm{L} \nonumber\\
	& = j^\tau_\mathrm{LR} + j^\tau_\mathrm{RL}
\end{align}
in terms of the left-right and right-left handed currents $j^\tau_\mathrm{LR}$ and $j^\tau_\mathrm{RL}$, respectively.

In three-quark system, heavy-light meson currents are recovered, e.\,g., for the choice $\tau = \widetilde\tau = (\lambda^4 + i\lambda^5)/2$ being a combination of Gell-Mann matrices.
Accordingly, we obtain
\begin{align}
	j_\mathrm{P}^{\widetilde\tau} & = i \bar \psi \gamma_5 {\widetilde\tau} \psi = i (\bar u,\bar d,\bar c) \gamma_5 \begingroup\renewcommand*{\arraystretch}{1.} \left( \begin{array}{ccc} 0&0&1 \\ 0&0&0 \\ 0&0&0 \end{array} \right) \left( \begin{array}{c} u \\ d \\ c \end{array} \right) \endgroup = i \bar u \gamma_5 c \nonumber\\
	&
	= i (\bar u_\mathrm{L} c_\mathrm{R} - \bar u_\mathrm{R} c_\mathrm{L})
\end{align}
and
\begin{align}
	j_\mathrm{S}^{\widetilde\tau} & = \psi {\widetilde\tau} \psi = \bar u c \nonumber\\
	&
	= \bar u_\mathrm{L} c_\mathrm{R} + \bar u_\mathrm{R} c_\mathrm{L} \, .
\end{align}

General chiral transformations restricted to the light parts of the left and right handed flavor vectors $\psi_\mathrm{L,R}$ read
\begin{align}
	\label{eq:chitransf}
	\psi_\mathrm{L,R} = \left( \begingroup\renewcommand*{\arraystretch}{1.} \begin{array}{c} u \\ d \\ c \end{array} \endgroup \right)_\mathrm{L,R} \quad\longrightarrow\quad \psi'_\mathrm{L,R} = e^{-i \Theta^a_\mathrm{L,R} \lambda^a/2} \psi_\mathrm{L,R}
\end{align}
with the rotation parameters $\Theta^a_\mathrm{L,R} = (\Theta^1_\mathrm{L,R},\Theta^2_\mathrm{L,R},\Theta^3_\mathrm{L,R},0,\ldots,0)$ and the Gell-Mann matrices $\lambda^a$.
Applying the $\mathrm{SU}(N_\mathrm{f}=2)$ finite chiral transformation formula employing the identity $e^{-i\Theta^a_C\sigma^a/2} = \cos\frac{|\Theta_C|}{2}-i\frac{\Theta_C^a\sigma^a}{|\Theta_C|}\sin\frac{|\Theta_C|}{2}$ with $|\Theta_C|=\sqrt{\sum_{a=1}^3(\Theta^a_C)^2}$ to the light-flavor components $\varphi = (u,d)^\mathrm{T}$ we can explicate the desired finite transformations
\begin{align}\label{eq:finChiTrafo}
	\psi_C & = \left( \begingroup\renewcommand*{\arraystretch}{1.} \begin{array}{c} \varphi_C \\ c_C \end{array} \endgroup \right)
	&&\longrightarrow
	&&\psi'_C = \left( \begingroup\renewcommand*{\arraystretch}{1.} \begin{array}{c} \varphi'_C \\ c_C \end{array} \endgroup \right) = \left( \begingroup\renewcommand*{\arraystretch}{1.} \begin{array}{c} \left[ \cos\frac{|\Theta_C|}{2} - i \frac{\Theta^a_C \sigma^a}{|\Theta_C|} \sin\frac{|\Theta_C|}{2} \right) \varphi_C \\ c_C \end{array} \endgroup \right] \, , \nonumber\\[0.5ex]
	\bar\psi_C & = \left( \bar\varphi_C , \bar c_C \right)
	&&\longrightarrow 
	&&\bar\psi'_C = \left( \bar\varphi'_C , \bar c_C \right) = \left( \bar\varphi_C \left[ \cos\frac{|\Theta_C|}{2} + i \frac{\Theta^a_C \sigma^a}{|\Theta_C|} \sin\frac{|\Theta_C|}{2} \right] , \bar c_C \right) \, ,
\end{align}
where $C$ is a common label for either $\mathrm{L}$ or $\mathrm{R}$, $\varphi_C=(u_C,d_C)^\mathrm{T}$, $\sigma^a$ are the Pauli matrices and $\Theta^a_C$ the three non-vanishing rotation parameters.
Employing the finite transformations restricted to the light part of the flavor vector $\psi$ we aim for a set of rotation parameters $\Theta^a_\mathrm{L,R}$ which transforms the pseudo-scalar into the scalar heavy-light current, i.\,e.\ $j_\mathrm{P}^{\widetilde\tau} \longrightarrow \left( j_\mathrm{P}^{\widetilde\tau} \right)' = j_\mathrm{S}^{\widetilde\tau} = \bar u_\mathrm{L} c_\mathrm{R} + \bar u_\mathrm{R} c_\mathrm{L}$:
\begin{align}
	\left( j_\mathrm{P}^{\widetilde\tau} \right)'
	& = i \left( \bar\psi'_\mathrm{L} \widetilde\tau \psi'_\mathrm{R} - \bar\psi'_\mathrm{R} \widetilde\tau \psi'_\mathrm{L} \right) \allowdisplaybreaks\nonumber\\[0.5ex]
	& = i
	\left( \bar u_\mathrm{L} , \bar d_\mathrm{L} , \bar c_\mathrm{L} \right)
	\left( \begingroup\renewcommand*{\arraystretch}{1.}
													  \begin{array}{ccc} \multicolumn{2}{c}{\multirow{2}{*}{$\displaystyle \left[ \cos\frac{|\Theta_\mathrm{L}|}{2} + i \frac{\Theta^a_\mathrm{L} \sigma^a}{|\Theta_\mathrm{L}|} \sin\frac{|\Theta_\mathrm{L}|}{2} \right]$}} & 0 \\
													  & & 0 \\
														0 & 0 & 1 \end{array} \endgroup \right)
	\left( \begingroup\renewcommand*{\arraystretch}{1.} \begin{array}{ccc} 0&0&1 \\ 0&0&0 \\ 0&0&0 \end{array} \endgroup \right) \nonumber\\
	& \quad\times \left( \begingroup\renewcommand*{\arraystretch}{1.}
														\begin{array}{ccc} \multicolumn{2}{c}{\multirow{2}{*}{$\displaystyle \left[ \cos\frac{|\Theta_\mathrm{R}|}{2} - i \frac{\Theta^a_\mathrm{R} \sigma^a}{|\Theta_\mathrm{R}|} \sin\frac{|\Theta_\mathrm{R}|}{2} \right]$}} & 0 \\
													  & & 0 \\
														0 & 0 & 1 \end{array} \endgroup \right)
	\left( \begingroup\renewcommand*{\arraystretch}{1.} \begin{array}{c} u_\mathrm{R} \\ d_\mathrm{R} \\ c_\mathrm{R} \end{array} \endgroup \right)
	- (\mathrm{L} \longleftrightarrow \mathrm{R}) \allowdisplaybreaks\nonumber\\[1.5ex]
	& = i
	\left( \bar u_\mathrm{L} \left[ \cos\frac{|\Theta_\mathrm{L}|}{2} + i \frac{\Theta^3_\mathrm{L}}{|\Theta_\mathrm{L}|} \sin\frac{|\Theta_\mathrm{L}|}{2} \right] c_\mathrm{R} + \bar d_\mathrm{L} \left[ i \frac{\Theta^1_\mathrm{L} + i \Theta^2_\mathrm{L}}{|\Theta_\mathrm{L}|} \sin\frac{|\Theta_\mathrm{L}|}{2}\right] c_\mathrm{R} \right) \nonumber\\
	& \quad- (\mathrm{L} \longleftrightarrow \mathrm{R}) \, .
\end{align}
Choosing
\begin{align}
	\label{eq:rotparam}
	\Theta^1_\mathrm{L} & = \Theta^2_\mathrm{L} = 0 \, , & \quad \Theta^3_\mathrm{L} & = (4k - 1)\pi \, , & \quad |\Theta_\mathrm{L}| & = \left| \Theta^3_\mathrm{L} \right| \, , && \nonumber\\
	\Theta^1_\mathrm{R} & = \Theta^2_\mathrm{R} = 0 \, , & \quad \Theta^3_\mathrm{R} & = (4k + 1)\pi \, , & \quad |\Theta_\mathrm{R}| & = \left| \Theta^3_\mathrm{R} \right|&&
\end{align}
with integer $k$ we obtain
\begin{align}
	\left( j_\mathrm{P}^{\widetilde\tau} \right)'
	& = i \left[ \bar u_\mathrm{L} \left( \cos\frac{\pi}{2} + i \frac{-\pi}{\pi} \sin\frac{\pi}{2} \right) c_\mathrm{R} + \bar d_\mathrm{L} \left( i \frac{0}{\pi} \sin\frac{\pi}{2} \right) c_\mathrm{R} \right] \nonumber\\
	& \quad - i \left[ \bar u_\mathrm{R} \left( \cos\frac{\pi}{2} + i \frac{\pi}{\pi} \sin\frac{\pi}{2} \right) c_\mathrm{L} + \bar d_\mathrm{R} \left( i \frac{0}{\pi} \sin\frac{\pi}{2} \right) c_\mathrm{R} \right] \nonumber\\
	& = \bar u_\mathrm{L} c_\mathrm{R} + \bar u_\mathrm{R} c_\mathrm{L} \nonumber\\
	& = j_\mathrm{S}^{\widetilde\tau} \, ,
\end{align}
where $k=0$ has been used, exemplarily.

The chiral transformation \eqref{eq:chitransf} specified by the rotation parameters \eqref{eq:rotparam} also exhibits the chirally odd nature of the chiral condensate
\begin{align}
	\langle \bar\varphi \varphi \rangle'
	& 
	= \langle \bar\varphi'_\mathrm{L} \varphi'_\mathrm{R} + \bar\varphi'_\mathrm{R} \varphi'_\mathrm{L} \rangle
	= \langle \bar\varphi_\mathrm{L} e^{i\pi\sigma^3} \varphi_\mathrm{R} \rangle + \langle \bar\varphi_\mathrm{R} e^{-i\pi\sigma^3} \varphi_\mathrm{L} \rangle
	= - \langle \bar\varphi_\mathrm{L} \varphi_\mathrm{R} + \bar\varphi_\mathrm{R} \varphi_\mathrm{L} \rangle 
	= - \langle \bar\varphi \varphi \rangle \, ,
\end{align}
i.\,e., as expected, it turns the chiral condensate into its negative.

\section[Temperature effects on the pseudo-scalar and scalar OPEs]{Temperature effects on the pseudo-scalar and scalar \glspl{OPE}}
\label{app:TdepOPE}

\begin{figure}[!t]
\settoheight{\imageheight}{\includegraphics[trim=0mm 0mm 0mm 10mm,clip,width=0.49\columnwidth]{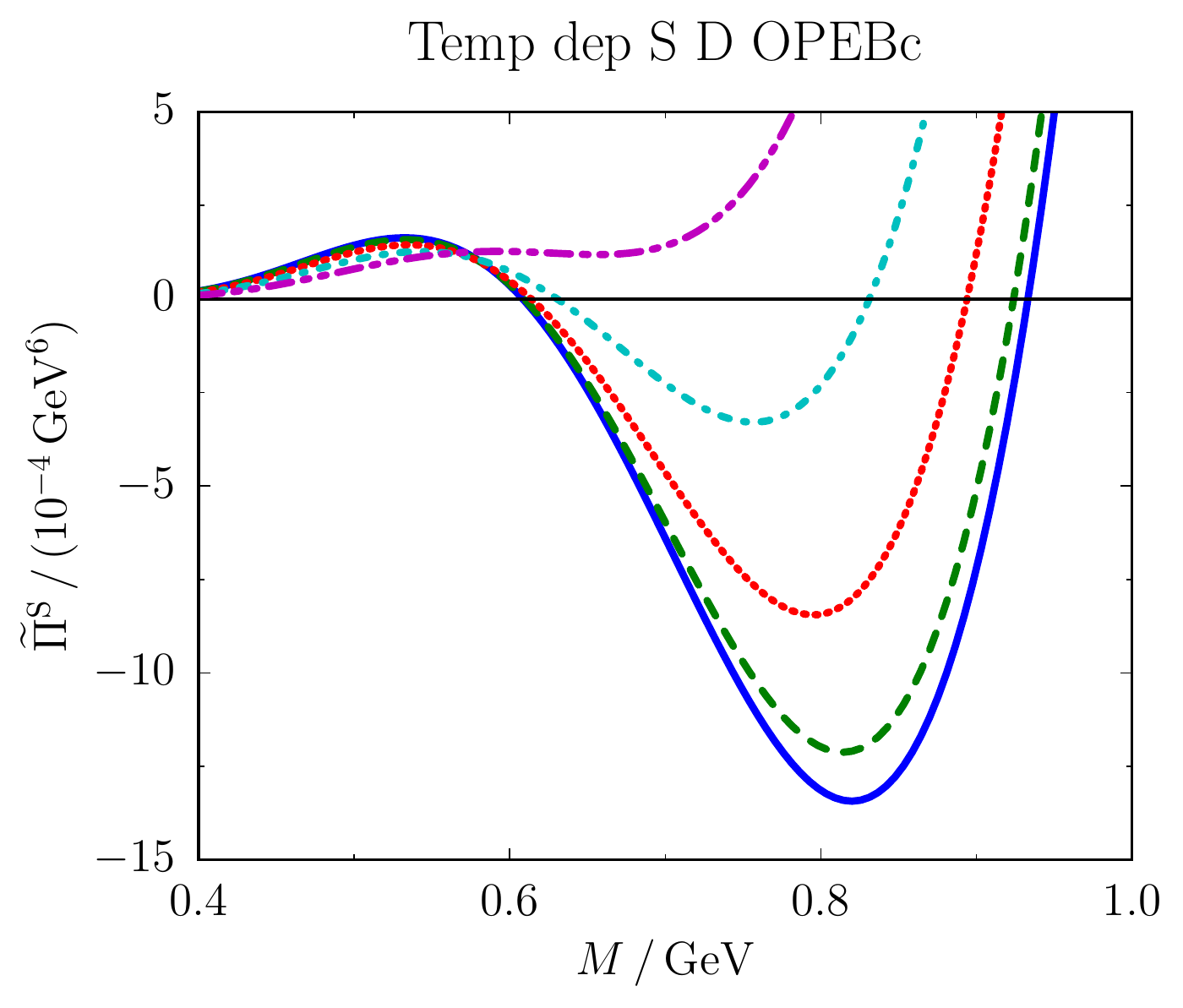}}
\includegraphics[trim=0mm 0mm 0mm 10mm,clip,height=\imageheight]{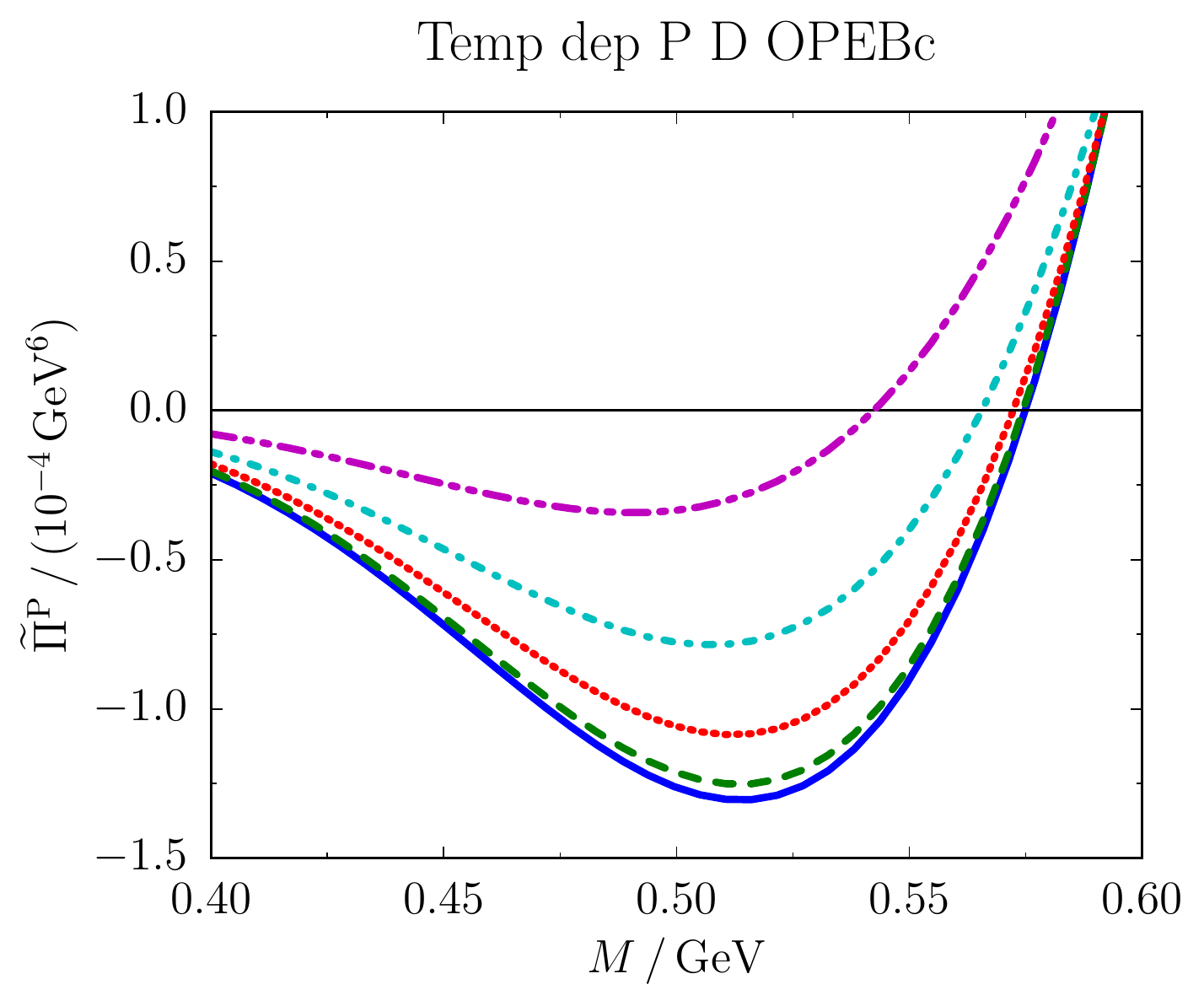}%
\hfill
\includegraphics[trim=0mm 0mm 0mm 10mm,clip,width=0.49\columnwidth]{fig10b.pdf}%
\caption[Comparison of temperature behavior of pseudo-scalar and scalar D meson \gls{OPE} Borel curves]{\gls{OPE} Borel curves of pseudo-scalar (left panel) and scalar (right panel) D mesons containing condensate contributions up to mass dimension 5 and with fixed continuum threshold parameters $s_0^\mathrm{P,S}=7\,\mathrm{GeV^2}$ at different temperatures: blue solid curve -- vacuum; green dashed, red dotted, cyan dot-dashed and magenta dot-dot-dashed curves are at $T=50$, 100, 150 and $200\,\mathrm{MeV}$, respectively.}%
\label{fig:OPEdrift}%
\end{figure}

The sizable shifts and vanishing of the poles of the mass Borel curve $m_\mathrm{S}(M)$ at higher temperatures can be understood from the scalar \gls{OPE} $\widetilde\Pi_\mathrm{S}(M)$ which drifts upwards for increasing temperatures featuring no zeros $M^\mathrm{S}_0$ above a particular temperature, cf.\ right panel in Fig.~\ref{fig:OPEdrift}.
In contrast, the temperature drift of the pseudo-scalar \gls{OPE} $\widetilde\Pi_\mathrm{S}(M)$, cf.\ left panel in Fig.~\ref{fig:OPEdrift}, does alter the location of its zero $M_0^\mathrm{P}$ on a smaller scale, i.\,e.\ $M_0^\mathrm{P}|_{T=0}-M_0^\mathrm{P}|_{T=150\,\mathrm{MeV}} \simeq 0.01\,\mathrm{GeV}$ with $M_0^\mathrm{S}|_{T=0}-M_0^\mathrm{S}|_{T=150\,\mathrm{MeV}} \simeq 0.1\,\mathrm{GeV}$.

\begin{figure}[!t]
\includegraphics[trim=0mm 0mm 0mm 10mm,clip,width=0.49\columnwidth]{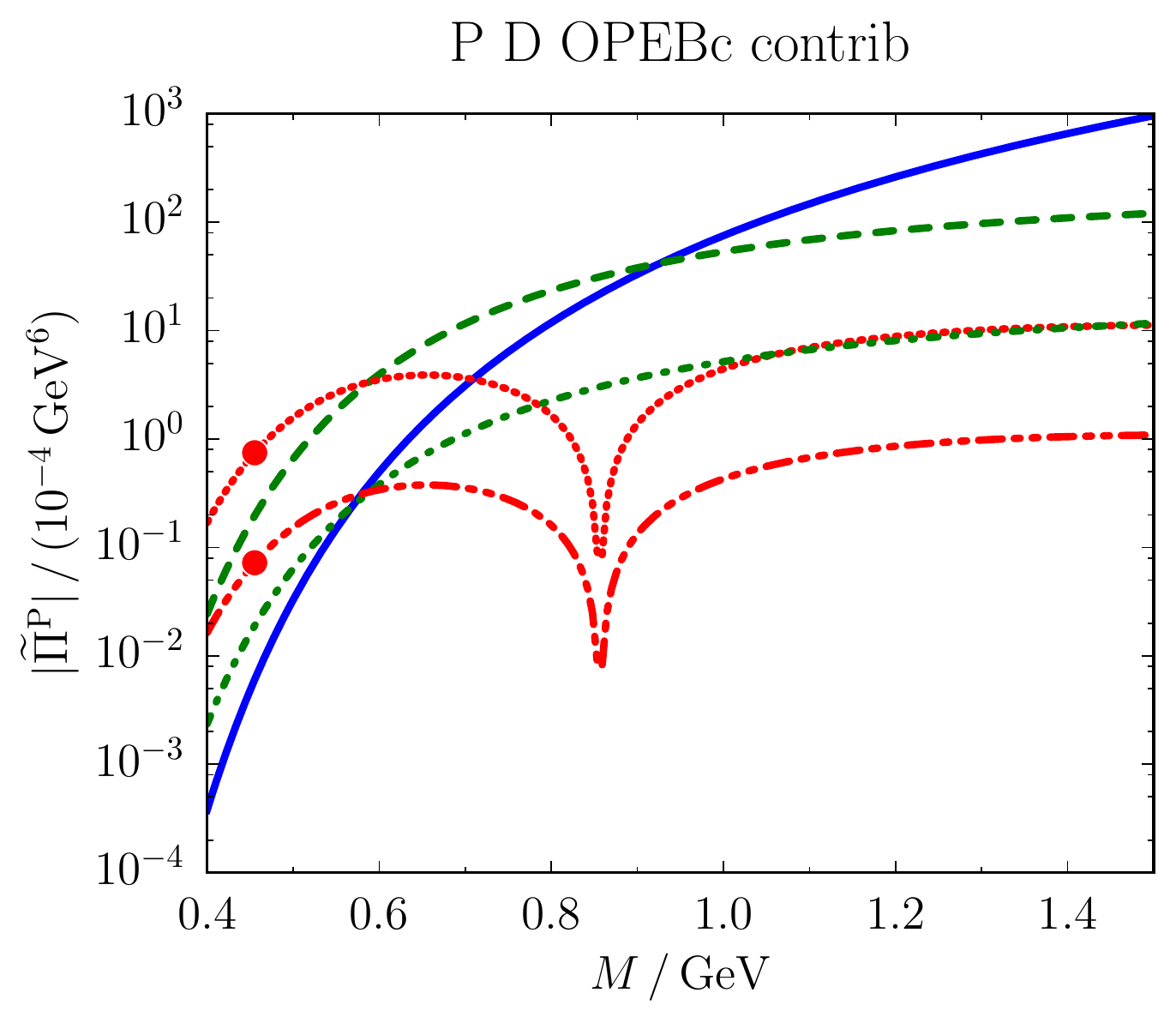}%
\hfill
\includegraphics[trim=0mm 0mm 0mm 10mm,clip,width=0.49\columnwidth]{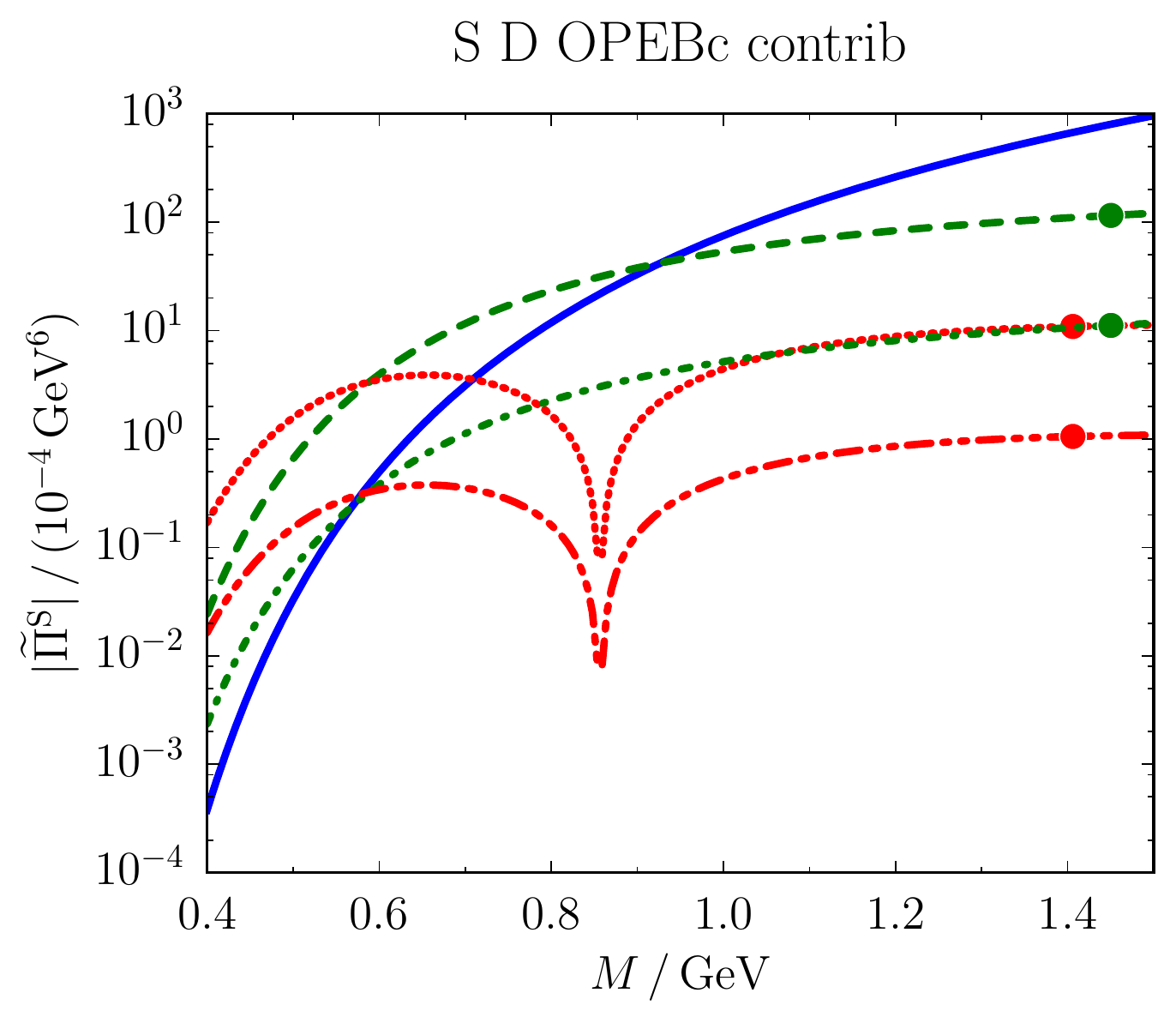}%
\caption[Temperature behavior of the predominant D meson \gls{OPE} contributions]{Modulus of the major \gls{OPE} contributions of pseudo-scalar (left panel) and scalar (right panel) D mesons with fixed continuum threshold parameters $s_0^\mathrm{P,S}=7\,\mathrm{GeV^2}$: the blue solid curve depicts the perturbative contribution, the green dashed and dot-dashed curves are the chiral condensates contributions in vacuum and at $T=250\,\mathrm{MeV}$, respectively, while the red dotted and dot-dot-dashed curves display the mixed condensate term in vacuum and at $T=250\,\mathrm{MeV}$, respectively. The branches with bullet markers originate from negative values.}%
\label{fig:OPEcontrib}%
\end{figure}

The vanishing and persistence of zeros of the pseudo-scalar and scalar \glspl{OPE} can also be understood if the major \gls{OPE} contributions at the relevant Borel mass ranges are considered.
In Fig.~\ref{fig:OPEcontrib} the 
\pagebreak
main contributions to the \gls{OPE} are depicted: the perturbative term, the chiral condensate term and the mixed quark-gluon condensate term.
The perturbative contribution remains the same while the condensate contributions decrease with increasing temperature.
In vacuum, the dominant contribution to the \gls{OPE} at high Borel mass $M$ is the perturbative term consecutively superseded by the chiral and mixed condensate term for lower values of $M$.
Due to the downshift of the condensate curves at high temperatures, e.\,g.\ $T=250\,\mathrm{MeV}$, the perturbative term as the dominating term of the OPE is directly superseded by the mixed condensate term for decreasing $M$.
Depending on the particular signs of the single condensate contributions this leads to different numbers of zeros of the \glspl{OPE} $\widetilde\Pi$, as exhibited in Fig.~\ref{fig:OPEdrift}.

In order to further investigate the poles of the mass Borel curves~\eqref{eq:ratioQSR} which originate from dividing by $\widetilde\Pi_X=0$ it is instructive to study $m_X(R_X)$ plots extracted from the sum rule $F$ and its derivative $F_1$, as depicted in Fig.~\ref{fig:crossings}, for values of $M$ in the vicinity of the zero of the corresponding \gls{OPE} $\widetilde\Pi_X$.
From $F = R_X e^{-m_X^2/M^2} - \widetilde\Pi_X(M)=0$ one expects vanishing residua $R_X$ or meson masses $m_X$ tending towards infinity for values of $M$ approaching $M_0^X$.
Indeed, in vacuum and with $s_0^X=7\,\mathrm{GeV}^2$, the $m_X(R_X)$ curve originating from $F$ produces a vertical section along $R_X=0$ followed by an approximately horizontal section which drifts upwards as $M$ approaches $M_0^X$.
The C-shape curves from $F_1$ remain unaffected.
For $M$ further apart from the pole, e.\,g.\ $|M-M_0^\mathrm{P}|>0.021\,\mathrm{GeV}$, the resulting curves of $F$ and $F_1$ intersect within a reasonable $R_X$-range, e.\,g.\ $R_\mathrm{P} < 2\,\mathrm{GeV}^6$, but in the close vicinity of $M_0^\mathrm{P}$ crossings tend to appear far above that regime indicating the singularity in the corresponding residuum Borel curve.
As this reasoning also applies to the temperature shifted poles, the vacuum as well as finite-$T$ mass and residuum Borel curves near the zeros of the \gls{OPE} have to be taken with care.

\section{Optimized QSR approach for decay constant extraction}
\label{app:luchaQSR}

For a pole $+$ continuum ansatz, the continuum threshold parameter $s_0$ is adjusted to reproduce the given meson mass parameter $m$ from the respective mass Borel curve $m(M)$ employing the flatness criterion.
Subsequently, the Borel averaged residuum $R$ is calculated from $\widetilde\Pi_X$ with the extracted continuum threshold parameter
.
The corresponding decay constant $f$ is readily obtained from Eq.~\eqref{eq:fRrel}.

In order to improve the flatness of the mass Borel curve within the Borel window one may introduce a Borel mass dependent continuum threshold parameter \cite{Lucha:2009et}
\begin{align}\label{eq:defLuchas}
	s_0(M) = \sum_{n=0}^{n_\mathrm{max}} \frac{s_{(n)}}{M^{2n}} \, ,
\end{align}
where the coefficients $s_{(n)}$ are chosen to minimize deviations of the mass Borel curve from the known actual meson mass.
This approach has been checked for potential toy models, where the spectral information of the lowest resonance as well as the \gls{OPE} are precisely known.
As an effective continuum threshold~\eqref{eq:defLuchas} produces more accurate results than a fixed continuum threshold parameter in these test cases \cite{Lucha:2009uy} one may infer that the Borel mass dependent $s_0$ reduces the contamination of the lowest resonance by continuum states, thus, rendering the semi-local quark-hadron duality $\int_{s_0}^\infty \rmd s \,e^{-s/M^2} \mathrm{Im}\Pi^\mathrm{cont}(s) \approx \int_{s_0}^\infty \rmd s \,e^{-s/M^2} \mathrm{Im}\Pi^\mathrm{pert}(s)$ exact if the fixed value of $s_0$ on the r.\,h.\,s.\ is substituted by the $M$-dependent one \cite{Lucha:2011zp}.
Due to the $M$-dependent continuum threshold parameter, further terms contribute to the derivative sum rule used to determine the mass Borel curve, i.\,e.\ for an original \gls{QSR} of the form
\begin{align}
	\int\limits_0^{s_0(M)} \rmd s \, e^{-s/M^2} \rho^\mathrm{res}(s) = \frac{1}{\pi} \int\limits_{m_Q^2}^{s_0(M)} \rmd s \, e^{-s/M^2} \mathrm{Im}\Pi^\mathrm{pert}(s) + \text{power corrections}
\end{align}
the derivative sum rule reads
\begin{align}
	& \int\limits_0^{s_0(M)} \rmd s \,s\, e^{-s/M^2} \rho^\mathrm{res}(s) + e^{-s_0(M)/M^2} \rho^\mathrm{res}\big(s_0(M)\big)\; \partial_{-M^{-2}} \Big( s_0(M) \Big) \nonumber\\[-0.8ex]
	& \qquad = \frac{1}{\pi} \int\limits_{m_Q^2}^{s_0(M)} \rmd s \,s\, e^{-s/M^2} \mathrm{Im}\Pi^\mathrm{pert}(s) + \frac{1}{\pi} e^{-s_0(M)/M^2} \mathrm{Im}\Pi^\mathrm{pert}\big(s_0(M)\big)\; \partial_{-M^{-2}} \Big( s_0(M) \Big) \nonumber\\[-0.3ex]
	& \qquad\phantom{=} + \partial_{-M^{-2}} \Big( \text{power corrections} \Big) \, .
\end{align}
While $\rho^\mathrm{res}\big(s_0(M)\big)=0$ for a pole ansatz $\delta(s-m^2)$, because we assume $m^2<s_0$, the continuum contribution merged with the perturbative term is altered if the derivative of the continuum threshold parameter w.\,r.\,t.\ $1/M^2$ does not vanish.

\baselineskip10pt

\bibliographystyle{aip}
\bibliography{lit}

\end{document}